\documentclass[12pt]{article} 

\newcommand{\be}{\begin{equation}}
\newcommand{\ee}{\end{equation}}
\newcommand{\bea}{\begin{eqnarray}}
\newcommand{\eea}{\end{eqnarray}}

\newcommand{\gs}{\ensuremath{g_s}} 
\newcommand{\ap}{\ensuremath{\alpha'}} 
\newcommand{\ls}{\ensuremath{l_s}} 

\def\p{\partial}
\def\pbar{\bar{\partial}}
\def\zbar{\bar{z}}
\def\para{{\scriptscriptstyle ||}}

\newcommand{\bS}{{\mathbf{S}}}

\newcommand{\bZ}{{\mathbf{Z}}}

\newcommand{\Gs}{\ensuremath{G_s}} 
\newcommand{\Go}{\ensuremath{G_o^2}} 
\newcommand{\ape}{\ensuremath{\alpha'_{e}}} 
\newcommand{\Ls}{\ensuremath{L_s}} 
\newcommand{\talpha}{\ensuremath{\tilde{\alpha}}}
\newcommand{\tbeta}{\ensuremath{\tilde{\beta}}}
\newcommand{\tgamma}{\ensuremath{\tilde{\gamma}}}

\begin{document}

\begin{titlepage}

\begin{flushright}
UUITP-11/00\\
USITP-00-18\\  
hep-th/0012183
\end{flushright}

\vspace{1cm}

\begin{center}
{\huge\bf Newtonian Gravitons and \\

\smallskip

D-brane Collective Coordinates\\

\smallskip

in Wound String Theory \\}

\end{center}
\vspace{3mm}

\begin{center}

{\large Ulf H.\ Danielsson,$^{\scriptstyle 1}$
Alberto G\"uijosa,$^{\scriptstyle 2}$
and
Mart\'\i n Kruczenski$^{\scriptstyle 1}$} \\

\vspace{5mm}

$^{1}$ Institutionen f\"or Teoretisk Fysik, Box 803, SE-751 08
Uppsala, Sweden

\vspace{3mm}

$^{2}$ Institute of Theoretical Physics, Box 6730, SE-113 85
Stockholm, Sweden

\vspace{5mm}

{\tt
ulf@teorfys.uu.se, alberto@physto.se, \\
martin.kruczenski@teorfys.uu.se \\
}

\end{center}

\vspace{5mm}

\begin{center}
{\large \bf Abstract}
\end{center}
\noindent

Recently it was shown
that NCOS theories are part of a ten-dimensional 
theory known as Non-relativistic Wound string theory. 
We clarify
the sense in which gravity is present in this theory.
We show
that Wound string theory contains exceptional unwound strings, including a 
graviton, which mediate the previously discovered
instantaneous long-range interactions, but 
are negligible as asymptotic states.  Unwound 
strings also provide the expected collective coordinates for the 
transverse D-branes in the theory. These and other results are shown to 
follow both from a direct analysis of the effect of the NCOS limit on 
the parent string theory, and from the worldsheet formalism developed 
by Gomis and Ooguri, about which we make 
some additional remarks.  We also devote some attention to 
supergravity duals, and in particular show that the open and closed 
strings of the theory are respectively
described by short and long strings on the 
supergravity side.

\vfill
\begin{flushleft}
December 2000
\end{flushleft}
\end{titlepage}
\newpage

\section{Introduction}
    
The emergence of noncommutativity in string 
theory \cite{cds,dh,sw} 
has recently been brought into sharper focus through the discovery 
of Noncommutative Open String (NCOS) \cite{sst2,ncos,om}
and Open Brane (OM/OD$p$) \cite{om,bbss2,harmark2} 
theories. These theories have generated interest not only because they 
display a certain form of 
noncommutativity between space and time, but also 
because they stand midway between field 
theories and conventional string/M-theory, since 
their fluctuation spectrum is believed not to include gravitons.

NCOS theory in $p+1$ dimensions ($p\le 5$) is obtained
by considering a D$p$-brane (or a stack of them)
in a low-energy limit $\ls\to 0$ where
the electric field on the brane is made to approach its critical 
value \cite{sst2,ncos}. 
These theories
were initially believed to contain only open strings, which
in particular means they do not include gravity.
In an important paper,
Klebanov and Maldacena \cite{km} discovered that, when
the direction of the electric field is compactified on a circle of 
radius $R$,
closed strings with strictly positive winding number
are also present in the theory. The graviton ($w=0$)
was still thought to be
absent, even though the spectrum does include
its $w>0$ cousins. 

Since (for finite $R$) closed strings can leave the brane, 
one is actually dealing with a $D$-dimensional
theory ($D=10$ for the superstring). 
It is then natural to wonder whether the theory makes sense 
even in the absence of the D$p$-brane(s), and recently
this question has been 
answered affirmatively \cite{wound,go}.
As explained in those works, 
starting with any one of the conventional string theories, it
is possible to single out a spatial direction (we will
take it to be $x^{1}$, and refer to it as the longitudinal direction),
compactify it on a circle of radius $R$, 
and consider a
limit where the coupling constant, string length, and (closed 
string) metric scale as
\be \label{woundlim}
\gs={\Gs\over\sqrt{\delta}}\to\infty,\quad  
\ls=\Ls\sqrt{\delta}\to 0, \quad
g_{\mu\nu}=(-1,1,\delta,\delta,\ldots), 
\mbox{ with } \Gs,\Ls,R\mbox{ fixed}.
\ee
The result is a consistent 
$D$-dimensional theory
characterized by the fact that all objects in it must 
carry strictly positive F-string winding along the longitudinal
direction, and consequently designated 
Wound String theory in \cite{wound}.
This theory was discovered 
independently by Gomis and Ooguri \cite{go},
who chose to refer to it 
as Non-relativistic (or Galilean) Closed String theory,
to emphasize the fact that its closed string spectrum has a 
non-relativistic form.  
The parameters $\Gs$ and $\Ls$ introduced in 
(\ref{woundlim}) are the effective coupling constant and string length
of the theory.\footnote{As was stressed in \cite{wound}, the coupling 
constant of the theory is really the dimensionful combination
$\Gs\Ls=\gs\ls$. The relation between
$\Gs,\Ls$ and the NCOS parameters $\Go,\ape$
is discussed in \cite{wound} and in Section \ref{longgosec} of the 
present paper.} 

States without any D-branes contain of course only closed strings 
(with $w>0$), 
but if there are D-branes in the spectrum of the parent 
string theory, the same will be true for the corresponding Wound string 
theory. To comply with the $w>0$ requirement, a D$p$-brane extended 
along the longitudinal direction must carry a near-critical electric 
field, thus giving rise to a setup which is exactly what is known as 
NCOS theory\footnote{Note that the NCOS name would not be 
appropriate for the full theory, not only because it contains closed 
strings, but also because space/time noncommutativity is not an 
intrinsic feature of the theory as a whole (e.g., there is 
no sign of it in closed string scattering amplitudes \cite{wound,go}).
Noncommutativity is a property only of the worldvolume of longitudinal 
D-branes in this theory.} \cite{wound,go}. Wound string theory
contains also transverse D-branes \cite{wound}.
 
Wound string theory is defined with a built-in compactification, but
one may of course consider
the decompactification limit
$R\to\infty$ as a special case. In this limit the wound strings 
decouple from the 
worldvolume theory on longitudinal D-branes, so the latter becomes the
original NCOS theory of \cite{sst2,ncos}, which was defined on a 
non-compact space. Note that this decoupling occurs not because the 
energy of the wound strings diverges in the limit (the 
D-branes carry F-string winding, so their energy diverges at the same 
rate),
but because the energy cost for a 
D-brane to emit a closed string into the bulk (i.e., the binding 
energy) is proportional to $R$.
In fact, when the theory is examined at finite temperature,
wound strings are seen to
play an important role even as $R\to\infty$.
Indeed, it was shown in \cite{ggkrw} that the Hagedorn transition 
in NCOS theories occurs when the temperature is large 
enough that the entropic contribution of these strings
to the free energy becomes larger than their binding energy,
resulting in their liberation from the D-brane.

As shown explicitly in \cite{wound,go} (see also \cite{hyun}),
longitudinal T-duality converts the limit (\ref{woundlim}) into 
the limit of discrete light-cone quantization (DLCQ) in the sense of 
\cite{seiberg,hp,sen}, so Wound string theory is T-dual to DLCQ 
string theory.\footnote{Notice this implies that the limit
$R\to\infty$ looks rather exotic from the DLCQ perspective: it
corresponds to shrinking the radius of the 
null circle to zero size.} 
It is also known to be
U-dual to various Wrapped (Galilean) Brane 
theories \cite{wound,go}, 
which contain the OM/OD$p$/NCYM theories as 
special classes of states \cite{wound}.
A Wrapped X$p$-brane theory is defined by starting with 
a parent string/M-theory 
compactified\footnote{The fact that the compactification
need not be toroidal was
first pointed out in \cite{go}. 
A simple but interesting example is M-theory compactified on 
$\bS^{1}\times\bS^{1}/\bZ_{2}$--- this yields a Wrapped M2-brane 
theory which is the eleven-dimensional lift of Wound Heterotic 
$E_{8}\times E_{8}$ string theory.} 
on a $p$-cycle, and then taking a limit 
analogous to (\ref{woundlim}), which truncates the spectrum down
to those objects carrying strictly positive 
X$p$-brane wrapping 
number on the $p$-cycle in question.
Because of their connection to DLCQ and Matrix theory 
\cite{bfss,susskind} on the one hand, and to the theories with 
noncommutativity on the other,
the Wound/Wrapped theories offer a novel and unified perspective
on several recurrent themes of recent years, and could thus
facilitate new developments.
With this motivation in mind, we will continue here
with the study of Wound 
string theory, making use of the two complementary approaches 
developed in \cite{wound} and \cite{go}. 

The central theme of this 
paper will be the presence of unwound closed strings in the theory, 
which are thus exceptions to the general $w>0$ rule. This exception 
was noted already in \cite{wound}, and will be elaborated upon in 
Section \ref{unwoundsec}. Strings with $w=0$ can survive the limit 
(\ref{woundlim}) only if they carry vanishing transverse momentum 
and have zero oscillator number; 
the graviton is among such exceptional states. 
Due to the $p_{\perp}=0$ restriction, the unwound strings
constitute a zero-measure set in phase 
space, and were therefore believed in \cite{wound} not to play any 
dynamical role in the theory. 

As will be explained in Sections 
\ref{polesec} and \ref{zerosec}, even though
their exceptional character indeed implies 
that they can be ignored as asymptotic states, the unwound strings 
do play one role: as the only massless objects in the 
theory, they are responsible for the 
Newtonian long-range interactions discovered in \cite{go}.
Whereas on-shell these strings cannot carry any transverse momentum 
(for otherwise their energy would diverge), off-shell 
their transverse momentum is unrestricted, allowing them to 
act as mediators of a long-range force. 
The metric rescaling in (\ref{woundlim}) effectively takes the 
transverse speed of light to infinity, so this force is transmitted 
instantaneously. Wound string 
theory is thus seen to contain Newtonian gravitons.

Section \ref{unwoundgosec} shows that this same conclusion 
follows from the worldsheet formalism of \cite{go},
and some additional remarks are made there
regarding that formalism. In 
particular, we note the existence of a free parameter which can 
be adjusted to simplify the form of the bosonic action, and
we write down the fermionic action. 

In Section \ref{dbranesec} we turn our attention to D-branes,
where the exceptional unwound strings find a second role:
open strings with $w=0$ and oscillator level $N=0$  
give rise to the expected collective coordinates for transverse
D-branes (as well as to Newtonian photons on the brane worldvolume). 
This is explained in Section \ref{transsec}, 
where the excitation spectrum for
transverse D-branes is shown to be of the same form as the
closed string spectrum.
In Section \ref{dbranegosec} this spectrum is reproduced
using the methods of \cite{go}, after having reviewed the
results of that paper for longitudinal D-branes, in order to clarify
their dependence on the parameters of the theory.

Section \ref{sugrasec} explains how some of the above properties of 
Wound string theory can be understood from the point of view of
the known supergravity backgrounds dual
to the longitudinal D$p$-branes of the theory (i.e., NCOS theory)
\cite{ncos,harmark1}. In particular, the point is made that
the supergravity dual describes not only 
the open strings attached to the branes, but also the closed $w>0$ 
strings that are free to move off of them. 

Our conclusions are summarized in
Section \ref{conclsec}, where we
include additional comments on the 
sense in which gravity is present in the theory,
and outline some of the directions for future work.

\section{The Physics of Unwound Strings} 
\label{unwoundsec}

\subsection{Long-range interactions from unwound strings}
\label{polesec}

As explained in \cite{wound,go}, all objects in Wound string theory carry
strictly positive F-string winding, and are consequently massive. We would
therefore expect interactions to be short-ranged, giving rise to potentials
which decrease exponentially with distance. It was discovered in \cite{go}
that this is actually not the case. To understand why, consider a process
where two wound strings scatter off each other in the parent string theory
(i.e., before taking the limit). For simplicity we will (as in \cite{go})
focus on `tachyons' (with positive winding number) in bosonic string theory.
For tachyons, the level matching condition $N_{L}-N_{R}=nw$ implies that $%
n=0 $ for all four vertices. The scattering amplitude is then
found to take the same form as the one for
unwound tachyons; this agreement is a trivial
consequence of T-duality. In particular, at tree level one obtains the
familiar Virasoro-Shapiro amplitude (see, e.g., \cite{polchinski}),
including a factor 
\begin{equation}
\Gamma (-1-{\frac{t}{4}}-{\frac{R^{2}}{4\ensuremath{\alpha'}}}%
(w_{1}+w_{3})^{2})~,
\end{equation}
where $t=-\ensuremath{\alpha'}(p_{1}+p_{3})^{2}$. One thus finds the
expected t-channel poles at energies such that 
$t=4N-(w_{1}+w_{3})^{2}R^{2}/\alpha'$ (where $N=-1,0,1,\ldots $), 
corresponding to the
usual spectrum for string theory on a circle. For the purpose of
extracting a potential, we restrict attention to the case of zero winding
transfer, $w_{3}=-w_{1}$. The poles are then at $t=-4,0,4,\ldots$ 
In the limit (\ref{woundlim})
we scale $\ensuremath{\alpha'}\rightarrow 0$, so all $t\neq 0$
poles move off to infinite energy. The $t=0$ pole suffers the same fate, 
\emph{unless} there is zero perpendicular-momentum transfer, 
$(\vec{k}_{1}+\vec{k}_{3})_{\perp}=0$ (where $\vec{k}_{\perp}$ is the 
transverse momentum in the coordinates where the metric scales
as in (\ref{woundlim})). 
Close to this pole the amplitude is proportional to 
$(\vec{k}_{1}+\vec{k}_{3})_{\perp}^{-2}$, 
which upon Fourier transformation 
gives rise to a
Newtonian potential \cite{go}.

This result is general: as discussed in \cite{wound}, tree-level closed
string scattering amplitudes in Wound string theory have the same form as
those in the parent theory, with the kinematic variables $s,t,u$ taking
their limiting values.\footnote{For some other amplitude calculations 
in Wound string theory see \cite{ncos,km,hk,fp}.} 
In all processes
one thus finds a t-channel pole which remains at finite
energy even when no winding number is exchanged. The presence of this pole
leads us to conclude that the spectrum of the Wound theory includes
exceptional closed string states with zero winding number. As seen above,
these states must carry vanishing transverse momentum, and are forced to have
oscillator levels $N_{L}=N_{R}=0$ (so they include the graviton). That these
states remain in the theory was noted already in \cite{wound}, although
their role as carriers of a Newtonian force was not recognized. 
Starting from the usual energy formula for a closed string, 
\begin{equation} \label{p0wB}
p_{0}=
\sqrt{\left({\frac{wR}{\alpha'}}\right) ^{2}
+\left( {\frac{n}{R}}\right) ^{2}+{\frac{|\vec{k}_{\perp }|^{2}}{\delta}}
+{\frac{2}{\alpha'}}(N_{L}+N_{R})}~, 
\end{equation}
it is seen that in the limit (\ref{woundlim})
a state with $w=0$ has a diverging
energy $p_{0}\propto\delta^{-1/2}$,
and is therefore removed from the spectrum \emph{unless} 
$\vec{k}_{\perp }=N_{L}=N_{R}=0$, in which case $p_{0}=|n|/R$.

Wound string theory is thus seen to contain gravitons, dilatons, and 
antisymmetricons
(plus massless Ramond-Ramond and fermionic states, if these 
were present in the parent theory). 
On-shell, these unwound strings are forced to have
vanishing transverse momenta, so they constitute a negligible zero-measure
set in the space of asymptotic states (we will return to this point in the
next subsection). Even so, the fact that they are the only massless ($w=0$)
states in the theory implies that, off-shell, they act as the sole mediators
of long-range interactions.

Notice that the fact that
these strings are unwound implies that their energy is finite
even in the decompactification limit $R\to\infty$. This means
in particular that they should show up
in the original NCOS one-loop scattering amplitudes. 
The nonplanar annulus amplitude for $2\to 2$ open string scattering
was considered in \cite{ncos}, and indeed, it can be seen from the 
equation following (4.4) on that paper that if $k_{\perp}=N=0$ then
there is a `pole' at finite $p_{0}=|p_{1}|$. As explained there, 
the pole appears integrated in momentum-space over the directions 
transverse to the D$p$-brane, and for $p<7$ there is 
no real singularity. 
The physical reason for this is that the brane, understood
as an object which is completely localized along the directions 
transverse to it, cannot emit a closed string with a definite 
transverse momentum (in this case, $k_{\perp}=0$). This 
conclusion was confirmed
in \cite{ggkrw}, where it was shown that
the cross-section for graviton production by the brane 
vanishes for all $p<7$. The statement 
that NCOS theories are decoupled from gravity
\cite{sst2,ncos,om,km,ggkrw}
is therefore seen to be related to the zero-measure character 
of the Newtonian gravitons (see the next subsection).  
At the same time, it is clear that, whether $R$ is finite or not,
the long-range interactions between D-branes (just 
like those between unwound strings) are 
controlled by off-shell strings with $w=0$.

The properties of the unwound strings also follow on very general grounds
from the nature of the limit (\ref{woundlim}). 
More specifically, note that the rescaling of the transverse metric
implies that the
relevant physics in the theory takes place on very small transverse scales.
As a consequence, 
the effective speed of light is taken to infinity, 
$c=\sqrt{|g_{\perp\perp}/g_{00}|}\sim 1/\sqrt{\delta}$, and
physics becomes non-relativistic. This is consistent with the limit 
(\ref{woundlim})
being equivalent (T-dual) to the DLCQ limit,
as was shown explicitly in \cite{wound,go} (see also \cite{hyun}). 
If we then consider the fate of a massless on-shell particle, obeying 
$E=ck$, as the speed of light is taken to infinity, we find that finite
energy indeed requires vanishing momentum in the limit.

As we have seen above,
the unwound strings are responsible for the instantaneous
gravitational force between the various objects of the theory. 
The force is instantaneous
because the speed of light is infinite. The carriers of the force are, as
usual, off-shell, which for finite $E$ and finite $k$ actually means that the
particles must be infinitely off-shell, i.e.,
$\Delta E\sim kc\sim 1/\sqrt{\delta }$. {}From the uncertainty
principle, the particles can then
exist only for a vanishingly short time 
$\Delta t\sim \hbar/\Delta E\sim \sqrt{\delta }$, 
which nevertheless allows them to reach any finite distance 
$s\sim c\Delta t\sim \hbar/k$.

Note that strings with $w=0$ are related by longitudinal T-duality to 
DLCQ strings with $p_{-}=0$. 
In the conventional treatment of DLCQ field theories \cite{my,pb}, 
states with vanishing longitudinal momentum
are not independent degrees of freedom; they satisfy a 
constraint which can in principle be used to eliminate them from the 
theory--- this is the infamous zero-mode problem
(for recent progress, see \cite{kmpv}).
When DLCQ is defined instead as a limit of compactification on a 
small spatial circle \cite{seiberg,hp}, as it is in our case,
the treatment of zero modes changes.
Their effect for
\emph{field} theories was analyzed in \cite{hp}, where they were shown 
to become strongly-coupled and thus complicate the analysis of the 
DLCQ limit (see also \cite{bilal3}). 
The situation in string/M-theory appears to be more 
benign:  unlike their field theory counterparts, 
perturbative string scattering amplitudes 
are well-defined in the DLCQ limit \cite{bilal,bilal2,uy}.
Through T-duality, this is true also for amplitudes in 
Wound string theory \cite{wound,go}.

\subsection{Unwound strings as asymptotic states}
\label{zerosec}

We would now like to show that unwound strings are physically 
irrelevant as asymptotic states: 
measurable quantities are always extracted
from scattering amplitudes by integrating over a phase space in which
unwound strings constitute a zero-measure set. To make this intuitive
argument more precise, consider as a concrete example the process
in the parent string theory through which a wound
string with energy, momentum and winding $(E,\vec{p},n,w)$
decays into two strings,
labelled 1 and 2, for which these quantities take the values $(E_{i},\vec{p}%
_{i},n_{i},w_{i})$, $i=1,2$. The decay rate takes the usual form 
\begin{equation}
\Gamma \propto {\frac{1}{R}}\sum_{n_{2}}\int {\frac{d{\vec{p}_{2}}}{%
(E-E_{2})E_{2}}}\delta (E-E_{1}-E_{2})\mathcal{A}(p,p_{2})~,  \label{decay}
\end{equation}
where the phase space integration is only over possible states for string 2;
the state of string 1 is determined by the conservation laws $\vec{p}_{1}=%
\vec{p}-\vec{p}_{2}$, $n_{1}=n-n_{2}$, $w_{1}=w-w_{2}$. The three-point
amplitude $\mathcal{A}(p,p_{2})$ is just a vertex involving the momenta and
polarization tensors for the three strings. Without loss of generality, we
can work in the reference frame where the decaying particle is at rest in
the transverse directions, $\vec{p}=0$.

Now, in terms of coordinates where the metric of the parent 
theory scales as in (\ref{woundlim}),
the coordinate momenta $\vec{k}_{1,2}$ are related to the
proper transverse momenta $\vec{p}_{1,2}$ through 
$\vec{p}=\vec{k}/\sqrt{\delta}$.
For any given value of $\delta $, we can use $\vec{k}_{2}$
instead of $\vec{p}_{2}$ as the integration variable in (\ref{decay}). The
Jacobian for this is $\delta ^{(D-2)/2}$, where $D$ is the total spacetime
dimension. 
Since the transition rate is given as the number of transitions
per unit volume and time, this Jacobian is precisely what is needed in order
to obtain a finite rate per unit coordinate volume, and we will 
therefore not write it explicitly. 

Consider first the case when string 2 is unwound: $w_{2}=N_{L,2}=N_{R,2}=0$.
Recalling the energy formula (\ref{p0wB}) we see that 
\begin{equation}
E_{2}=\sqrt{\left( {\frac{n}{R}}\right) ^{2}+{\frac{|\vec{k}_{2}|^{2}}{%
\delta }}}  \label{E2}
\end{equation}
and 
\begin{equation}
E-E_{1}=-{\frac{\ensuremath{L_s}^{2}}{2wR}}|\vec{k}_{2}|^{2}+{\frac{N-N_{1}}{%
wR}}-{\frac{|\vec{k}_{2}|}{\sqrt{\delta }}}+{\mathcal{O}}(\delta )~,
\label{E-E1}
\end{equation}
where we have let $N=N_{L}+N_{R}$. The delta-function in (\ref{decay}) thus
forces 
\begin{equation}
|\vec{k}_{2}|={\frac{(N-N_{1})}{wR}}\sqrt{\delta }+{\mathcal{O}}(\delta
^{3/2}),  \label{k2u}
\end{equation}
which through (\ref{E2}) implies that $E_{2}\sim {\mathcal{O}}(1)$. Using
the delta-function to dispose of the radial integral over $|\vec{k}_{2}|$ in
(\ref{decay}), we conclude that 
\begin{equation}
\Gamma _{w=0}\propto \delta ^{(D-2)/2}\left. \mathcal{A}\right| _{|\vec{k}%
_{2}|\sim {\mathcal{O}}(\sqrt{\delta })}  \label{decayu}
\end{equation}

Consider now the case when string 2 is wound, $w_{2}>0$. It is easy to see
that the energy delta-function in (\ref{decay}) then constrains $|\vec{k}%
_{2}|\sim {\mathcal{O}}(1)$, and therefore 
\begin{equation}
\Gamma _{w>0}\propto \left. \mathcal{A}\right| _{|\vec{k}_{2}|\sim {\mathcal{%
O}}(1)}  \label{decayw}
\end{equation}
In both cases $\mathcal{A}$ is a function of the transverse momenta $\vec{p}=%
\vec{k}/\sqrt{\delta }$ and the (finite) energies of the strings, which
so we conclude that in the limit of interest $\Gamma_{w=0}$ is
vanishingly small compared to $\Gamma_{w>0}$. 

A nice way to understand the result is to make an analogy with
bremsstrahlung. The energy loss for an
accelerated charged particle is given by 
\[
\frac{dW}{dt}\sim \frac{Q^{2}a^{2}}{c^{3}},
\]
where $a$ is the acceleration and $c$ is the speed of light. $Q^{2}$ (the
charge squared) is defined as having dimensions of
energy times length, so that
the potential energy for two charges is of the form 
\[
V=\frac{Q_{1}Q_{2}}{r}.
\]
We should now consider the non-relativistic limit, with the speed of light
going to infinity. While doing that we need to make sure that $Q$ is finite
in order for the strength of the electric forces to remain unaffected. In
this limit the bremsstrahlung goes away, showing that the energy loss is a
relativistic effect. This is of course connected with the fact that
fields are relativistic constructions. 
It is only in relativity, with a finite speed of
light, where fields are needed in order to carry a force--- fields
which also might serve as an energy dump as in the case of bremsstrahlung.
In a direct-action theory there is never any asymptotic state associated
with a carrier of the force, since there is always a recipient at the other
end of the line. As we have seen, the story is completely analogous in the
case of the unwound strings.

\subsection{Worldsheet perspective}
\label{unwoundgosec}

Wound string theory is defined as the specific limit (\ref{woundlim}) of a
standard string theory, so all of its properties can be deduced by focusing
on the corresponding aspect of the parent theory and studying the effect of
the limit. This is the approach adopted in \cite{sst2,ncos,km,wound} and in
the preceding subsections of this paper. A complementary approach, developed
by Gomis and Ooguri \cite{go}, is to take the limit once and for all at the
level of the worldsheet action. This has the advantage of producing a finite
worldsheet Lagrangian which serves as a more explicit definition of the
theory \cite{go}. Also, for actual calculations, using the resulting
worldsheet rules will in general be more convenient than taking the limit in
each case separately. On the other hand, in this approach the relation to
the parent theory is not transparent, and it is therefore important to make
sure that the formalism correctly captures all properties of the theory. 
We will now examine this question in relation to the unwound
strings discussed in the previous subsections, which, in fact, did not appear
in the analysis of \cite{go}.
We will find that the treatment of $w=0$ strings requires special care.
Along the way we will make some additional remarks regarding the 
Gomis-Ooguri formalism.

In the approach of \cite{go}, 
the bosonic part of
the usual string action in the presence of a $B_{01}$-field is first 
rewritten in the form\footnote{%
In comparing with \cite{go}, note that $B_{\mbox{here}}=2\pi %
\ensuremath{\alpha'}B_{\mbox{there}}$.} 
\[
S=\int {\frac{d^{2}z}{2\pi }}\left\{ \beta \bar{\partial}\gamma +%
\ensuremath{\tilde{\beta}}\partial 
\ensuremath{\tilde{\gamma}}-\frac{2(\ls/\Ls)^{2}}{1+B}\beta 
\ensuremath{\tilde{\beta}}+\frac{1-B}{2(\ls/\Ls)^{2}}
\partial \gamma \bar{\partial}%
\ensuremath{\tilde{\gamma}}+{\frac{1}{\ensuremath{L_s}^{2}}}\partial X^{i}%
\bar{\partial}X^{i}\right\} 
\]
where $i=2,\ldots ,D-1$, 
\begin{equation}
\ensuremath{L_s}\gamma \equiv \ensuremath{X^{+}}\equiv X^{0}+X^{1},\qquad %
\ensuremath{L_s}\ensuremath{\tilde{\gamma}}\equiv \ensuremath{X^{-}}\equiv
-X^{0}+X^{1},  \label{gammas}
\end{equation}
and $\beta ,\ensuremath{\tilde{\beta}}$ are Lagrange multipliers. 
The limit (\ref{woundlim}) is then taken
while simultaneously making the
$B$-field approach its critical value according to 
\begin{equation}
B\equiv B_{01}=1-\lambda \left( {\frac{\ensuremath{l_s}}{\ensuremath{L_s}}}%
\right)^{2},  \label{lambda}
\end{equation}
where we have included a free parameter $\lambda $. 
Keeping 
the leading and subleading terms in $(\ls/\Ls)^{2}$,
the result is
\begin{equation}
S=\int {\frac{d^{2}z}{2\pi }}\left\{ \beta \bar{\partial}\gamma +%
\ensuremath{\tilde{\beta}}\partial \ensuremath{\tilde{\gamma}}-
\left({\ls\over\Ls}\right)^{2}\beta \ensuremath{\tilde{\beta}}+%
\frac{\lambda }{2}\partial \gamma \bar{\partial}\ensuremath{\tilde{\gamma}}+{%
\frac{1}{\ensuremath{L_s}^{2}}}\partial 
X^{i}\bar{\partial}X^{i}\right\}~. 
\label{ourgoaction}
\end{equation}
Gomis and Ooguri \cite{go} then 
proceeded by putting $(\ls/\Ls)^{2}=0$ identically, thereby arriving 
at
\be \label{goaction}
S= \int {d^{2}z\over 2\pi} \left\{\beta\pbar\gamma + \tbeta\p\tgamma
   + {\lambda\over 2}\p\gamma\pbar\tgamma
   +{1\over\Ls^{2}}\p X^{i}\pbar X^{i} \right\}~.
\ee
The Lagrange multipliers then constrain $X^{+}$ and 
$X^{-}$ to be respectively analytic and 
antianalytic.\footnote{That this is the effect of the limit 
was pointed out already in \cite{sst2}.} 
As we will argue, however,
dropping the $\beta \ensuremath{\tilde{\beta}}$
term in (\ref{ourgoaction}) is strictly speaking correct only if 
$w\neq 0$. 

Let us now explain the significance of the parameter $\lambda$.
As explained in \cite{wound}, as far as closed strings are concerned, the
role of the $B$-field is merely to remove a divergent contribution to the
energy arising from the winding term $|w|R/\ensuremath{\alpha'}$. As always,
when subtracting this infinity one has the option of leaving behind a finite
term, and this freedom is parametrized by $\lambda $. The simplest choice is 
$\lambda =0$, which corresponds to not leaving any such term behind \cite
{wound}. The $\beta $-$\gamma $ action (\ref{goaction})
is then exactly that of a system of
commuting ghosts. For open strings associated with longitudinal D-branes the
situation is more subtle. At first sight, it would seem like $\lambda =0$ is
in that case not allowed; the authors of \cite{go} chose instead $\lambda
=1/2$, to conform with the usual NCOS convention \cite{sst2,ncos,km}. 
In Section \ref{longgosec} we will amplify the discussion of \cite{wound} in
this regard, emphasizing that the NCOS conventions are no longer useful now
that they are understood to refer only to a particular subsector of the full
Wound string theory. It will become clear there that the open string
spectrum is in fact independent of $\lambda $, so we can consistently set $%
\lambda =0$. In the meantime $\lambda $ is left arbitrary to keep track of
its effect on the expressions to follow.

Before proceeding with the review of \cite{go}, and the extension of their
results to $w=0$, let us note that the fermionic part of the action can be
easily put into a form analogous to the $\beta $-$\gamma $ system 
(\ref{goaction}). 
In the
presence of a background $B$-field, the action for the left-moving fermions
reads \cite{pum} 
\begin{equation}
S_{\psi }=\int {\frac{d^{2}z}{4\pi }}(g_{\mu \nu }+B_{\nu \mu })\psi ^{\mu }%
\bar{\partial}\psi ^{\nu }~.  \label{psiaction}
\end{equation}
For the transverse part of the action, the property $\int d\theta \,\exp
(-A\theta )=A\int d\theta \,\exp (-\theta )$ can be used to bring $g_{ii}=(%
\ensuremath{l_s}/\ensuremath{L_s})^{2}\rightarrow 0$ out of the fermionic
path integrals, where it can be absorbed through a rescaling of the overall
normalization factor. 
The limit (\ref{woundlim})$+$(\ref{lambda}) can then be taken without any
difficulty, yielding 
\begin{equation} \label{psigoaction}
S_{\psi }=\int {\frac{d^{2}z}{2\pi }}\left\{ b\bar{\partial}c+{\frac{1}{2}}%
\psi ^{i}\bar{\partial}\psi ^{i}\right\} ~,
\end{equation}
where we have defined $b=(-\psi ^{0}+\psi ^{1})/\sqrt{2}$, $c=(\psi
^{0}+\psi ^{1})/\sqrt{2}$. So, just like $X^{0,1}$ reduce in the limit to a
system of commuting ghosts (with conformal weights $h_{\beta }=1$, $%
h_{\gamma }=0$) \cite{go}, we see that $\psi ^{0,1}$ are equivalent to a
system of anticommuting ghosts
(with weights $h_{b}=h_{c}=1/2$). 
For simplicity, in the rest of the paper we will concentrate on the bosonic
part of the system.

Let us now proceed with the general analysis by writing down the equations
of motion: 
\begin{eqnarray}
\bar{\partial}\gamma  &=&\left({\ls\over\Ls}\right)^{2}%
\ensuremath{\tilde{\beta}},
\qquad \bar{\partial}\beta +\frac{\lambda }{2}%
\partial \overline{\partial }\widetilde{\gamma }=0, \\
\partial \ensuremath{\tilde{\gamma}} &=&\left({\ls\over\Ls}\right)^{2}\beta,
\qquad \partial \widetilde{\beta }+\frac{%
\lambda }{2}\partial \overline{\partial }\gamma =0,  \nonumber
\end{eqnarray}
Note how the presence of the $\beta\tilde{\beta}$-term
provides antianalytic contributions to $\gamma$, and analytic contributions
to $\tgamma$. The mode expansions are 
\begin{eqnarray} \label{premodeexp}
\beta (z) &=&\sum_{n=-\infty }^{\infty }\beta _{n}z^{-n-1}~,\qquad %
\ensuremath{\tilde{\beta}}(\bar{z})=\sum_{n=-\infty }^{\infty }%
\ensuremath{\tilde{\beta}}_{n}\bar{z}^{-n-1}~,  \\
\gamma (z,\bar{z}) &=&\left[ 
+i\left(\frac{wR}{\ensuremath{L_s}}%
\right)
+ {\ls^{2}\over\Ls^{2}}\ensuremath{\tilde{\beta}}_{0}\right]\log z
+\sum_{n=-\infty }^{\infty }\gamma _{n}z^{-n}+
{\ls^{2}\over\Ls^{2}}\ensuremath{\tilde{\beta}}%
_{0}\log \bar{z}-{\ls^{2}\over\Ls^{2}}%
\sum_{n\neq 0}{\frac{\ensuremath{\tilde{\beta}}_{n}}{n}}\bar{z}^{-n}~, 
\nonumber \\
\ensuremath{\tilde{\gamma}}(z,\bar{z}) &=&\left[
-i\left( \frac{wR}{\ensuremath{L_s}}\right)
+{\ls^{2}\over\Ls^{2}}\beta _{0}\right] \log \bar{z}
+\sum_{n=-\infty }^{\infty }\ensuremath{\tilde{%
\gamma}}_{n}\bar{z}^{-n}+{\ls^{2}\over\Ls^{2}}%
\beta _{0}\log z-{\ls^{2}\over\Ls^{2}}%
\sum_{n\neq 0}{\frac{\beta _{n}}{n}}z^{-n}~.  \nonumber
\end{eqnarray}
where $R$ is the radius of the longitudinal direction. (If this direction is
not compact we must restrict attention to the states with $w=0$.) The OPE's
imply that the only non-zero commutators are 
\begin{equation}
\lbrack \gamma _{n},\beta _{m}]=\delta _{n+m},\qquad \lbrack %
\ensuremath{\tilde{\gamma}}_{n},\ensuremath{\tilde{\beta}}_{m}]=\delta
_{n+m}.  \label{comm}
\end{equation}
The contribution from the directions 01 (`parallel' to $B$) to the
energy-momentum tensor is 
\begin{equation}
T^{{\scriptscriptstyle||}}(z)=-:\beta \partial \gamma :,\qquad T^{{%
\scriptscriptstyle||}}(\bar{z})=-:\ensuremath{\tilde{\beta}}\bar{\partial}%
\ensuremath{\tilde{\gamma}}:,  \label{Tpara}
\end{equation}
and the corresponding Virasoro modes are 
\begin{eqnarray} \label{Lpara}
L_{n}^{{\scriptscriptstyle||}} &=&\left[
-i\left( \frac{wR}{\ensuremath{L_s}}\right)
-\left({\ls\over\Ls}\right)^{2}\ensuremath{\tilde{\beta}}_{0}\right]
\beta _{n}+\sum_{m}m:\beta _{n-m}\gamma
_{m}:~,   \\
\tilde{L}_{n}^{{\scriptscriptstyle||}} &=&\left[ 
+i\left( \frac{wR}{\ensuremath{L_s}}\right) 
-\left({\ls\over\Ls}\right)^{2}\ensuremath{\tilde{\beta}}_{0}\right] 
\ensuremath{\tilde{\beta}}_{n}+\sum_{m}m:%
\ensuremath{\tilde{\beta}}_{n-m}\ensuremath{\tilde{\gamma}}_{m}:~.  \nonumber
\end{eqnarray}

Assume for now that $w>0$. Terms of order $(\ls/\Ls)^{2}$ are then
subleading in all expressions, and can therefore be dropped, as 
was done in \cite{go}. One then has, in particular, 
\begin{equation}
L_{0}^{{\scriptscriptstyle||}}=-i\beta _{0}\left( \frac{wR}{\ensuremath{L_s}}%
\right) +N_{{\scriptscriptstyle||}}~,\qquad \ensuremath{\tilde{L}}_{0}^{{%
\scriptscriptstyle||}}=+i\ensuremath{\tilde{\beta}}_{0}\left( \frac{wR}{%
\ensuremath{L_s}}\right) +\ensuremath{\tilde{N}}_{{\scriptscriptstyle||}}~.
\label{L0para}
\end{equation}
{}From (\ref{goaction}), the momenta conjugate to $\gamma $ and $%
\ensuremath{\tilde{\gamma}}$ are
\begin{eqnarray}
\Pi _{\gamma } &\equiv &i{\frac{\partial \mathcal{L}}{\partial \dot{\gamma}}}%
=\ensuremath{L_s}\Pi _{+}={\frac{i}{2\pi }}\left( z\beta +{\frac{\lambda }{2}%
}\bar{z}\bar{\partial}\ensuremath{\tilde{\gamma}}\right) ~,  \label{pigamma}
\\
\Pi _{\ensuremath{\tilde{\gamma}}} &\equiv &i{\frac{\partial \mathcal{L}}{%
\partial \dot{\ensuremath{\tilde{\gamma}}}}}=\ensuremath{L_s}\Pi _{-}={\frac{%
i}{2\pi }}\left( \bar{z}\ensuremath{\tilde{\beta}}+{\frac{\lambda }{2}}%
z\partial \gamma \right) ~.  \nonumber
\end{eqnarray}
(The dot denotes differentiation with respect to $\sigma _{2}=\ln |z|$.) The
zero mode piece of these equations reads 
\begin{eqnarray} \label{cmomenta}
\ensuremath{L_s}\ensuremath{p_{+}} &\equiv &\ensuremath{L_s}{1\over 2}%
(+p_{0}+p_{1})=i\beta _{0}+{\frac{\lambda }{2}}\left( \frac{wR}{%
\ensuremath{L_s}}\right) ~,  \\
\ensuremath{L_s}\ensuremath{p_{-}} &\equiv &\ensuremath{L_s}{1\over 2}%
(-p_{0}+p_{1})=i\ensuremath{\tilde{\beta}}_{0}-{\frac{\lambda }{2}}\left( 
\frac{wR}{\ensuremath{L_s}}\right) ~,  \nonumber
\end{eqnarray}
from which it follows that \cite{go}
\begin{eqnarray}
p_{0} &=&{\frac{i(\beta _{0}-\ensuremath{\tilde{\beta}}_{0})}{%
\ensuremath{L_s}}}+\lambda \left( \frac{wR}{\ensuremath{L_s}^{2}}\right) ~,
\label{pbeta} \\
p_{1} &=&{\frac{i(\beta _{0}+\ensuremath{\tilde{\beta}}_{0})}{%
\ensuremath{L_s}}}~.  \nonumber
\end{eqnarray}
If the longitudinal direction is compact, then of course $p_{1}=n/R$.

Let us now determine the spectrum. The perpendicular directions give rise to
the usual Virasoro modes, including 
\begin{equation}
L_{0}^{\perp }={\frac{\ensuremath{L_s}^{2}}{4}}p_{\perp }^{2}+N_{\perp
},\qquad \ensuremath{\tilde{L}}_{0}^{\perp }={\frac{\ensuremath{L_s}^{2}}{4}}%
p_{\perp }^{2}+\ensuremath{\tilde{N}}_{\perp }.  \label{L0perp}
\end{equation}
For convenience, we use transverse number operators $N_{\perp },%
\ensuremath{\tilde{N}}_{\perp }$ which are shifted by $-1$ with respect to
the usual definition,\footnote{%
The shift by -1 refers to the bosonic string. For the superstring we would
shift by -1/2 in the NS sector, and by 0 in the R sector.} so that the
physical state conditions read 
\begin{equation} \label{phys}
L_{n}\equiv L_{n}^{{\scriptscriptstyle||}}+L_{n}^{\perp }=0,\qquad %
\ensuremath{\tilde{L}}_{n}\equiv \ensuremath{\tilde{L}}_{n}^{{%
\scriptscriptstyle||}}+\ensuremath{\tilde{L}}_{n}^{\perp }=0\qquad \mbox{for
all}\quad n\geq 0. 
\end{equation}
The possible eigenvalues of $N_{\perp },\ensuremath{\tilde{N}}_{\perp }$ are
then $-1,0,1,\ldots $ The level-matching condition $L_{0}=%
\ensuremath{\tilde{L}}_{0}$ implies\footnote{%
Due to our conventions, for the NS-R sector of the superstrings there would
actually be an additional constant in (\ref{levelmatch}).} 
\begin{equation}
N_{{\scriptscriptstyle||}}+N_{\perp }-\ensuremath{\tilde{N}}_{{%
\scriptscriptstyle||}}-\ensuremath{\tilde{N}}_{\perp }=nw~.
\label{levelmatch}
\end{equation}
The spectrum follows from the requirement $L_{0}=\ensuremath{\tilde{L}}_{0}=0
$. As shown in \cite{go}, through (\ref{L0para}), (\ref{pbeta}), and (\ref
{L0perp}) these two conditions imply that for $w\neq 0$, 
\begin{equation}
p_{0}=\lambda {\frac{wR}{\ensuremath{L_s}^{2}}}+{\frac{\ensuremath{L_s}%
^{2}p_{\perp }^{2}}{2wR}}+{\frac{N+\ensuremath{\tilde{N}}}{wR}}~,
\label{p0wound}
\end{equation}
where $N=N_{{\scriptscriptstyle||}}+N_{\perp }$, $\ensuremath{\tilde{N}}=%
\ensuremath{\tilde{N}}_{{\scriptscriptstyle||}}+\ensuremath{\tilde{N}}%
_{\perp }$. This spectrum
was originally derived in \cite{km}. 
Incidentally, notice that the finite negative energy assigned by 
(\ref{p0wound}) 
to a state with $w<0$ is an artifact of the formalism; in reality
such states have energies which are positive and infinitely larger than
those of the positively wound states \cite{wound}.

Having reviewed the results of \cite{go}, 
let us now consider the unwound
states, $w=0$. 
We focus on the left-movers, but the story is analogous for
the right-movers. Using (\ref{L0para}) and (\ref{L0perp}), 
we learn that in order to satisfy 
$L_{0}=\ensuremath{\tilde{L}}_{0}=0$ we must demand that
$(\ensuremath{L_s}p_{\perp})^{2}=-4N=-4\ensuremath{\tilde{N}}$. 
Other than the tachyon, 
the only possibility is $p_{\perp}=N=\ensuremath{\tilde{N}}=0$, i.e.,
`gravitons' with zero perpendicular momentum. These are precisely the
unwound strings discussed in the previous subsections. \emph{A priori}, 
they can be polarized along the transverse directions 
($N_{{\scriptscriptstyle||}}=0$, $N_{\perp }=0$) 
or along the parallel directions 
($N_{{\scriptscriptstyle||}}=1$, $N_{\perp}=-1$). 
To ascertain this we must
remember to enforce (\ref{phys}) not only for $n=0$, but also
for all $n>0$. The expectation would then be 
that any time-like polarization
would be removed by the constraints, 
while a longitudinal polarization would be gauge. 
Unfortunately, one discovers that the $L_{1}$ constraint, as given
by (\ref{Lpara}) with $w=0$ and $(\ls/\Ls)^{2}=0$, requires only that
$\beta_{0}\gamma_{1}=0$. 
If we in addition have $\beta_{0}=0$ (i.e., 
$p_{-}=0$ but in general $p_{+}\neq 0$), 
the constraint fails to remove both
the $\gamma _{-1}$ and the $\beta _{-1}$ states. Furthermore, since
the $\beta _{0}\gamma_{-1}$ term
in $L_{-1}$ similarly disappears, 
neither the $\gamma _{-1}$
nor the $\beta _{-1}$ are spurious for $\beta_{0}=0$. The result is
therefore a situation where all polarizations are physical, and there
are undesired negative-norm states. 

However, it should be clear that the cause of the trouble is that we have
erroneously generalized results derived for $w>0$ to the case of $w=0$. 
The point is that for $w=0$, the \emph{leading} piece of
the expression inside the square brackets in 
(\ref{Lpara}) is of order $(\ls/\Ls)^{2}$. 
For arbitrarily small but finite $(\ls/\Ls)^{2}$
this term should not be discarded. 
With it one find expressions for $L_{\pm 1}$ that are
adequate for removing unwanted states. If $p_{-}=0$, for instance,
one finds that the state involving $\gamma_{-1}$ is 
unphysical, while the one involving $\beta_{-1}$ is spurious.
We thus conclude that only 
unwound states with transverse polarizations are
truly physical--- these are the expected Newtonian gravitons.

As we have seen in Section \ref{zerosec}, on-shell it 
is really only states with strictly 
positive winding that are relevant in the theory,
so it is of interest to determine which states are physical for 
$w>0$. For this purpose,
it is actually quite
useful to fall back on more familiar language,
rewriting $\gamma,\beta,\tgamma,\tbeta$ in terms of $X^{+},X^{-}$.  
Comparing the mode expansions of $\Ls\gamma$ and $X^{+}$ 
($\Ls\tgamma$ and $X^{-}$), it is easy to 
see that 
\be \label{gtoa}
\alpha^{+}_{n}=-i\sqrt{2}n\gamma_{n}, \qquad 
\talpha^{-}_{n}=-i\sqrt{2}n\tgamma_{n} \qquad \forall\quad n\neq 0.
\ee
At the classical level, the effect of the limit (\ref{woundlim}) is to
remove the antianalytic (analytic) piece of $X^{+}$ ($X^{-}$), but from the
commutators (\ref{comm})
we see that in fact 
\be \label{btoa}
\talpha^{+}_{n}=i\sqrt{2}\tbeta_{n}, \qquad 
\alpha^{-}_{n}=i\sqrt{2}\beta_{n} \qquad \forall\quad n\neq 0.
\ee
Similarly, for the zero modes one can deduce that
\bea \label{bgtoaz}
\alpha^{+}_{0}=-\sqrt{2}\left({wR\over\Ls}\right), &\qquad& 
\talpha^{+}_{0}=i\sqrt{2}\tbeta_{0}=\sqrt{2}\Ls p_{-},\qquad \\
\talpha^{-}_{0}=+\sqrt{2}\left({wR\over\Ls}\right), &\qquad& 
\alpha^{-}_{0}=i\sqrt{2}\beta_{0}=\sqrt{2}\Ls p_{+}, \nonumber
\eea
where we have used (\ref{cmomenta}), setting $\lambda=0$ for simplicity.
The Virasoro modes for the entire system then have the standard form,
\be \label{La}
L_{n}={1\over 2}\sum_{m}g_{MN}:\alpha_{m}^{M}\alpha_{n-m}^{N}:~,
\ee
where $M=(+,-,i)$ and $g_{+-}=g_{-+}=1/2,g_{ii}=1$.
It is convenient to define the usual left- and right-moving momenta 
$p^{M}_{\mbox{\scriptsize L}}=(\sqrt{2}/\Ls)\alpha^{M}_{0}$, 
$p^{M}_{\mbox{\scriptsize R}}=(\sqrt{2}/\Ls)\talpha^{M}_{0}$, i.e.,
\bea \label{lrmomenta}
p_{\mbox{\scriptsize L}M}&=&(p_{+},-{wR\over\Ls^{2}},p_{i})~, \\
p_{\mbox{\scriptsize R}M}&=&(+{wR\over\Ls^{2}},p_{-},p_{i})~. \nonumber
\eea
{}From (\ref{La}) we then have in particular
\bea \label{Laz}
L_{0}&=&{\Ls^{2}\over 4}g_{MN}
p^{M}_{\mbox{\scriptsize L}}
p^{M}_{\mbox{\scriptsize R}}
+N_{\para}+N_{\perp}, \\
\tilde{L}_{0}&=&
{\Ls^{2}\over 4}g_{MN}
p^{M}_{\mbox{\scriptsize L}}
p^{M}_{\mbox{\scriptsize R}}
    +\tilde{N}_{\para}+\tilde{N}_{\perp}. \nonumber
\eea
We thus see that all expressions have the usual form, except for one
peculiarity:
if from (\ref{lrmomenta}) we try to read off left- and right-moving 
momenta along directions $01$ in the `obvious' way, 
these would not have the standard form 
$p^{0}_{\mbox{\scriptsize L}}=p^{0}_{\mbox{\scriptsize R}}=p^{0}$,
$p^{1}_{\mbox{\scriptsize L,R}}=p^{1}\pm wR/\Ls^{2}$; 
in particular, the left and right components of
$p_{0}$ would not be equal. This is then the main modification that 
the analysis of \cite{go} brings to light.
We should also note that, if one wishes to consider the 
analog of the above dictionary for the states with zero winding,
it is again important to retain terms of order $(\ls/\Ls)^{2}$ in 
place of the terms involving $w$ 
in (\ref{bgtoaz}) and (\ref{lrmomenta}).

Given the formal agreement with the familiar expressions,
the Virasoro constraints can be imposed
level by level in the usual manner. 
In particular, one finds that the polarizations of, e.g.,
gravitons, are required to be transverse to the momenta as 
given by (\ref{lrmomenta}). 
{}From this OCQ analysis one concludes that negative-norm states are 
removed from the physical spectrum in a way consistent with the general
no-ghost theorem for Wound string theory, proven in \cite{go}
(for $w>0$) by means of BRST methods. 
Notice however that, contrary to what 
the authors of \cite{go} appear to indicate, for $w>0$
there are physical states in the theory polarized along the `parallel' 
directions (i.e., having $N_{\para}\neq 0$).

The non-relativistic character of Wound string theory,  
apparent from  the form of the wound 
string spectrum (\ref{p0wound}) and from
the existence of Newtonian gravitons, is ultimately
expressed by the fact
that the action (\ref{goaction}) is invariant under the 
Galilean group\footnote{This 
invariance is in accord 
with the T-duality relation to DLCQ string theory \cite{wound,go}.} 
\cite{go}. The invariance under translations and transverse
rotations is evident. A Galilean boost should have the form 
\begin{equation}
X^{i}\rightarrow X^{i}+{\frac{v^{i}}{2}}\ensuremath{L_s}(\gamma -%
\ensuremath{\tilde{\gamma}}),  \label{galilean}
\end{equation}
with $\gamma ,\ensuremath{\tilde{\gamma}}$ (and therefore $X^{0},X^{1}$)
invariant. It is easy to check that if the remaining fields transform
according to 
\begin{eqnarray}
\beta  &\rightarrow &\beta -{\frac{v^{i}}{\ensuremath{L_s}}}\partial X^{i}-{%
\frac{\vec{v}^{2}}{4}}\partial (\gamma -\ensuremath{\tilde{\gamma}}),
\label{galilean2} \\
\ensuremath{\tilde{\beta}} &\rightarrow &\ensuremath{\tilde{\beta}}+{\frac{%
v^{i}}{\ensuremath{L_s}}}\partial X^{i}+{\frac{\vec{v}^{2}}{4}}\partial
(\gamma -\ensuremath{\tilde{\gamma}}),  \nonumber
\end{eqnarray}
then the Lagrangian in (\ref{goaction}) changes only by a total derivative.
The effect on the modes can be most easily summarized for all $n$ in terms
of $\alpha _{n}^{M },\ensuremath{\tilde{\alpha}}_{n}^{M }$: 
\begin{eqnarray}
&{}&x^{i}\rightarrow x^{i}+v^{i}x^{0}  \label{galmodes} \\
&{}&\alpha _{n}^{i}\rightarrow \alpha _{n}^{i}+{\frac{v^{i}}{2}}\alpha
_{n}^{+},\qquad \qquad \quad \;\;\,\ensuremath{\tilde{\alpha}}%
_{n}^{i}\rightarrow \ensuremath{\tilde{\alpha}}_{n}^{i}-{\frac{v^{i}}{2}}%
\ensuremath{\tilde{\alpha}}_{n}^{-},  \nonumber \\
&{}&\alpha _{n}^{-}\rightarrow \alpha _{n}^{-}-v^{i}\alpha _{n}^{i}-{\frac{%
\vec{v}^{2}}{4}}\alpha _{n}^{+},\qquad \ensuremath{\tilde{\alpha}}%
_{n}^{+}\rightarrow \ensuremath{\tilde{\alpha}}_{n}^{+}+v^{i}%
\ensuremath{\tilde{\alpha}}_{n}^{i}-{\frac{\vec{v}^{2}}{4}}%
\ensuremath{\tilde{\alpha}}_{n}^{-},  \nonumber
\end{eqnarray}
with $\alpha _{n}^{+}$ and $\ensuremath{\tilde{\alpha}}_{n}^{-}$ invariant.
Specializing to $n=0$, these relations can be seen to imply that the momenta
transform according to 
\[
p_{0}\rightarrow p_{0}-v^{i}p_{i}+{\frac{\vec{v}^{2}}{2}}\left( {\frac{wR}{%
\ensuremath{L_s}^{2}}}\right) ,\qquad p_{1}\rightarrow p_{1},\qquad
p_{i}\rightarrow p_{i}-v^{i}\left( {\frac{wR}{\ensuremath{L_s}^{2}}}\right) .
\]
This is as expected for a Galilean boost, with $\mu=wR/\ensuremath{L_s}^{2}$
playing the role of mass--- a fact which was already inferred in \cite
{wound,go} from the form of the wound string spectrum (\ref{p0wound}).
Notice this explains why the $p_{i}=0$ restriction for the unwound states is
not incompatible with transverse boosts: these states have vanishing
Newtonian mass ($\mu =0$), so their momentum is Galilean-invariant.

\section{D-branes (Mostly Transverse)}
\label{dbranesec}

\subsection{Spectrum and collective coordinates}
\label{transsec}

As explained in \cite{wound,go},
longitudinal D-branes in Wound string theory comply with the 
requirement of carrying strictly positive F-string winding by 
supporting
a near-critical electric field on their worldvolume, thus giving
rise to a setup which is precisely what is known as NCOS 
theory \cite{sst2,ncos}. The spectrum
of open strings ending on such branes has the standard form 
(although as explained in Section \ref{longgosec}, the relevant open 
string metric is non-standard). 
D-branes which are transverse to the compact $x^{1}$ 
direction are also present in the theory \cite{wound}; their 
excitation spectrum 
is non-standard because of the $w>0$ requirement. Naively,
this restriction appears to indicate that the fluctuation spectrum
for transverse
branes does not include the usual collective 
coordinates \cite{wound}, 
since the latter 
are associated with $w=0$ strings. This conclusion seems rather
peculiar,
since the objects in question are expected to break translational 
invariance even after taking the limit.
To clarify the situation,
let us now explicitly
work out the spectrum for open strings ending on a transverse D-brane. 

The 
bosonic worldsheet action in the parent theory (prior to the limit)
reads
\be \label{Xaction}
S={1\over 2\pi\ap}\int d^{2}z \left[
\eta_{ab}\p X^{a} \pbar X^{b}
+B(\p X^{0}\pbar X^{1}-\p X^{1}\pbar X^{0})
+{\ap\over\Ls^{2}}\p X^{i}\pbar X^{i} \right]~,
\ee
where $a,b=0,1$ and $i=2,\ldots,D-1$. The equations of motion 
$\p\pbar X^{\mu}=0$ hold if we enforce boundary conditions at 
$z=\zbar$ such that 
\bea \label{fullbc}
\delta X^{0}\left[(\p-\pbar)X^{0}+B(\p+\pbar)X^{1}\right]&=&0, \\
\delta X^{1}\left[(\p-\pbar)X^{1}+B(\p+\pbar)X^{0}\right]&=&0,\nonumber\\
\delta X^{i}\left[(\p-\pbar)X^{i}\right]&=&0.\nonumber
\eea
For a \emph{transverse} D$p$-brane we expect to be able to choose
\be \label{transfullbc}
(\p-\pbar)X^{0}=0, \qquad \delta X^{1}=0,
\ee
and Neumann (Dirichlet) boundary conditions along $p$ ($D-p-2$) of the 
transverse directions.  Notice these conditions
are all indeed compatible with 
(\ref{fullbc}). 
We can then follow the standard procedure\footnote{For longitudinal 
D-branes things are actually not this simple--- 
see, e.g., \cite{aasj1}-\cite{ch2}. As is well-known by now, the end
result in that case can be succinctly summarized by introducing
an effective open 
string metric, coupling constant, 
and non-commutativity parameter \cite{sw}.}: 
we write
$X^{a}(z,\zbar)=X^{a}_{L}(z)+X^{a}_{R}(\zbar)$, and implement the 
usual doubling trick $X^{0}_{L}(\zbar)=X^{0}_{R}(\zbar)$,
$X^{1}_{L}(\zbar)=-X^{1}_{R}(\zbar)$ (and similarly for $X^{i}$).
The mode expansions are
\bea \label{transfullmodeexp}
X^{0}(z,\zbar)&=&x^{0}-i\sqrt{\ap\over 2}\alpha^{0}_{0}\log z\zbar
 +i\sqrt{\ap\over 2}\sum_{m\neq 0}{\alpha^{0}_{m}\over m}
 (z^{-m}+\zbar^{-m})~, \\
X^{1}(z,\zbar)&=&x^{1}-iwR\log{z\over\zbar}
 +i\sqrt{\ap\over 2}\sum_{m\neq 0}{\alpha^{1}_{m}\over m}
 (z^{-m}-\zbar^{-m})~, \nonumber
 \eea
and the expressions for the conjugate momenta, 
\bea \label{transfullcmom}
\Pi_{0}\equiv i{\p\mathcal{L}\over\p\dot{X}^{0}}&=&
  {i\over 2\pi\ap}\left[ z\p X_{0}+\zbar \pbar X_{0}
  -Bz\p X_{1}+B\zbar\pbar X_{1}\right]~, \\
\Pi_{1}\equiv i{\p\mathcal{L}\over\p\dot{X}^{1}}&=&
  {i\over 2\pi\ap}\left[ z\p X_{1}+\zbar \pbar X_{1}
  -Bz\p X_{0}+B\zbar\pbar X_{0}\right]~, \nonumber
\eea
imply that the corresponding
zero modes are related through
\be \label{transfullp0p1}
p_{0}=-{1\over\sqrt{2\ap}}\alpha^{0}_{0}-{wRB\over\ap},
\qquad p_{1}=0.
\ee
We should note
that the \emph{total} longitudinal momentum does not vanish, 
due to a net contribution from the non-zero modes:
\be \label{P1}
P_{1}\equiv\int_{0}^{\pi} d\sigma^{1}\,\Pi_{1}
=-{i\over\pi\sqrt{2\ap}}\sum_{m\neq  0}{e^{-m\sigma^{2}}\over m}
  (\alpha^{1}_{m}+B\alpha^{0}_{m})~.
\ee
(The center of mass coordinate $\bar{x}^{1}\equiv 
(1/\pi)\int d\sigma^{1}\, X^{1}$ 
receives a similar time-dependent contribution 
from the non-zero modes.) This is just a reflection of the fact that 
the brane breaks translational invariance along $x^{1}$, so the  
total longitudinal
momentum is not conserved (this is true even if $B=0$).
Notice on the other hand that the time average of $P_{1}$
(with respect to $\tau=-i\sigma^{2}$) vanishes--- just as it should, 
since the string cannot wander away from the brane. 
The time-dependence of $P_{1}$ should not be cause for distraction:
it is $p_{1}$, the zero mode of $\Pi_{1}$, that is canonically 
conjugate to $x^{1}$.

It follows from the above that
\be \label{transfullL0para}
L^{\para}_{0}={1\over 2}\sum_{m}\eta_{ab}:\alpha^{a}_{-m}\alpha^{b}_{m}:
 = -\ap (p_{0})^{2}- 2p_{0}wRB + 
 {(wR)^{2}\over\ap}(1-B^{2})+N_{\para}~,
\ee
which together with 
\be \label{transL0perp}
L^{\perp}_{0}=\Ls^{2} p_{\perp}^{2}+N_{\perp}
\ee
implies that the spectrum is identical to that of closed 
strings (with $p_{1}=0$),
\be \label{transfullp0}
p_{0}=-{BwR\over\ap}+\sqrt{\left({wR\over\ap}\right)^{2}
      +{\Ls^{2}p_{\perp}^{2}\over\ap}+{N_{\para}+N_{\perp}\over\ap}}~.
\ee
The effect of the limit (\ref{woundlim})$+$(\ref{lambda})
on this spectrum was determined in 
\cite{km,wound}: 
states with $w>0$ will have a finite energy
\be \label{p0trans}
p_{0}=\lambda{wR\over \Ls^{2}}+{\Ls^{2}p_{\perp}^{2}\over 2wR}
    +{N_{\para}+N_{\perp}\over 2wR}~,
\ee
whereas states with $w=0$ will have a finite energy only if 
$p_{\perp}=N_{\para}+N_{\perp}=0$, in which case
$p_{0}=0$. 

The positively wound states were anticipated in \cite{wound}; they are 
the mechanism through which the transverse brane can comply with the 
requirement of carrying positive F-string winding. The 
unwound strings are an exception to this requirement, just like the 
Newtonian gravitons discussed in Section \ref{unwoundsec}. 
The exception is important here since it resolves the puzzle discussed 
at the beginning of this section: the $w=0$ strings are present only 
at oscillator level $N=0$, so these are precisely the modes giving rise
to the standard D-brane gauge field and collective coordinates.
Notice that the constraint $p_{\perp}=0$ applies to all 
transverse directions, including those along the D$p$-brane. For a 
D-particle this constraint is automatically satisfied. 
For $p\ge 1$,
it implies that the `photons' on the brane, just like the gravitons in 
the bulk, are Newtonian: they propagate at infinite speed, and mediate 
instantaneous interactions between the charges associated with
string endpoints. Like the gravitons, they are negligible as 
asymptotic states.

It is also interesting to consider strings connecting two parallel 
transverse D-branes. If the branes are separated by a 
distance $L$ along the longitudinal direction, 
the $X^{0},X^{1}$ mode expansions for a string extending from 
one brane to the other are as in (\ref{transfullmodeexp}), with 
$wR$ replaced by $L/2\pi$. These strings can thus be regarded as a 
particular case of the wound open strings considered above, with 
fractional winding number $w\equiv L/(2\pi R)$. In particular, their energy 
spectrum in 
the limit (\ref{woundlim})$+$(\ref{lambda})
is given by (\ref{p0trans}). 
So in contrast with the rest of the 
objects in Wound string theory, whose Newtonian
masses are integer multiples of 
$R/\Ls^{2}$, the mass
of these strings is an arbitrary 
(positive) real number, $\mu=L/(2\pi\Ls^{2})$, which can 
remain finite in the decompactification limit $R\to\infty$.
Notice that $\mu\to 0$ when the branes approach one another, 
so again the energy of these strings remains finite only
if $p_{\perp}=N=0$. 

Before closing this section, let us address the question of whether 
the open string sector  analyzed above,
describing the excitations of a transverse 
D-brane,
can interact consistently with the closed string sector of 
Wound string theory. For the case of a longitudinal D-brane, (i.e., 
NCOS theory,) consistency was proven in \cite{km} by showing that a 
non-planar one-loop amplitude describing the scattering
of open strings attached to the brane has poles at energies which 
correspond precisely to the closed string spectrum (\ref{p0wound}). 
For the case of a transverse D-brane
it turns out to be easier to establish consistency--- the argument goes 
as follows.
Consider the non-planar annulus diagram in the parent string theory 
(before taking the limit). Notice first that the boundary 
conditions (\ref{transfullbc}) do not involve the $B$-field, so the 
propagator on the annulus is the standard one \cite{acny}. This implies 
that the annulus amplitude (see e.g., Eqs. (3.1)-(3.4) in \cite{ys}) 
depends on $B$ only through its effect on the masses of the 
open strings in the loop, 
which are summed over in the usual partition function. 
{}From the analysis in the present subsection, we know that this effect is 
extremely simple: as seen in (\ref{transfullp0}), the presence of the 
$B$ field merely effects an energy subtraction proportional to the winding.
For any finite value of $\delta$, the amplitude in question will 
clearly have poles at the energies given by the
closed string spectrum 
(\ref{p0wB}), where $B$ enters through the same energy subtraction.
Taking the limit (\ref{woundlim})$+$(\ref{lambda}), then, we obtain an 
amplitude whose poles necessarily
coincide with the closed string spectrum 
(\ref{p0wound}). In short, the point is that in this aspect, as in 
all others, the consistency of Wound string theory is guaranteed by 
the consistency of its parent theory.

\subsection{Worldsheet perspective}
\label{dbranegosec}

It is interesting to ask how transverse D-branes
are seen in the Gomis-Ooguri  
formalism \cite{go}. 
To answer this question,
one must consider
the worldsheet action (\ref{goaction}) in the presence of a boundary.
The equations of motion are then seen to imply the boundary conditions
\cite{go}
\be \label{gobc}
\delta\gamma(\beta-{\lambda\over 2}\pbar\tgamma)
+\delta\tgamma(-\tbeta+{\lambda\over 2}\p\gamma)=0~.
\ee
We will consider the theory formulated
on the upper-half plane; the boundary is then at $z=\zbar$. 
There should be at least two ways
of satisfying the above condition, 
corresponding to longitudinal and transverse D-branes. The former type 
of D-brane was shown in \cite{go} to lead to the standard NCOS 
setup \cite{sst2,ncos}. 
Let us first review that case, to clarify the dependence of the 
open string spectrum on the parameters of the theory.

\subsubsection{Longitudinal D-branes}
\label{longgosec}

For D-branes extended along the 
longitudinal direction it is clear that neither
$\gamma$ nor $\tgamma$ should satisfy 
Dirichlet boundary conditions, so we must demand that \cite{go}
\be \label{longbc1}
\beta|_{z=\zbar}={\lambda\over 2} \pbar\tgamma|_{z=\zbar}, \qquad
\tbeta|_{z=\zbar}={\lambda\over 2} \p\gamma|_{z=\zbar}~.
\ee
Before proceeding further it is important to realize that
for this type of D-brane 
it is not enough to specify the value of the 
background $B_{01}$-field.  
A longitudinal D-brane 
placed in such a field describes a bound state of a D-brane and a 
definite number  
$N$ of  fundamental strings. Since the total
F-string winding number $W$ is conserved by interactions, 
one can consistently
restrict attention to a sector of the theory with a particular value 
of $W$, but included in this sector are states where the D-brane 
carries different winding numbers \cite{wound}. In other words,
each D-brane can carry a different
electric field $E\equiv 2\pi\ap F_{01}$ 
on its worldvolume, whose value must be specified.

The combination $E+B$ determines the number $N$ of 
F-strings bound to the D$p$-brane through the flux
quantization condition
(see, e.g., \cite{ncos}) 
\be \label{pquantization}
N\gs=\left({r_{2}\cdot\cdot\cdot r_{p}\over 
\ls^{p-1}}\right)\sqrt{-g_{(p+1)}}
{|g^{00}|g^{11}(E+B)\over\sqrt{1-|g^{00}|g^{11}(E+B)^{2}}}~,
\ee
where $g_{(p+1)}$ is the determinant of the induced metric on the brane.
In the limit (\ref{woundlim}), 
this condition can be seen to imply
\be \label{quant}
E+B=1-{1\over 2\nu^{2}\Gs^{2}}\left({\ls\over\Ls}\right)^{2}, \qquad
\nu\equiv {N\Ls^{p-1}\over r_{2}\cdot\cdot\cdot r_{p}}~.
\ee
The description is physically most transparent in the gauge $B=0$; it is 
then evident that the value of $E$ (or equivalently, the number 
density $\nu$) should be specified separately 
for each brane \cite{wound}. In this description, the infinite 
contribution of F-string winding to the energy of each object of the 
theory must be subtracted by hand. Alternatively,
as explained in \cite{wound}, 
the subtraction can be implemented by gauging $E$ away 
from the brane and into the bulk, in the form of a $B$-field. 
The point we are emphasizing here is that 
with this single operation it is not possible to set $E=0$ on all 
possible longitudinal D-branes in the theory. 
After the 
gauge transformation $B$ is given by (\ref{lambda}),
so the electric field that remains on the D-brane worldvolume is  
$E=[\lambda-1/(2\nu^{2}\Gs^{2})](\ls/\Ls)^{2}$.
This (dimensionless) field vanishes in the limit, but it 
leaves behind a finite boundary action
\be \label{Eact}
\int d\tau\,A_{a}\p_{\tau}X^{a}
  ={1\over 2\pi\ap}\int d\tau\,E X^{0}\p_{\tau}X^{1}
  ={1\over 2\pi\Ls^{2}}\int d\tau\, \left[\lambda-{1\over 
  2\nu^{2}\Gs^{2}}\right] X^{0}\p_{\tau}X^{1}~.
\ee
When this is added to the bulk worldsheet action 
(\ref{goaction}) the boundary conditions
are modified: instead of (\ref{longbc1}) one must enforce
\be \label{longbc}
\beta|_{z=\zbar}={1\over 4\nu^{2}\Gs^{2}} \pbar\tgamma|_{z=\zbar}, \qquad
\tbeta|_{z=\zbar}={1\over 4\nu^{2}\Gs^{2}} \p\gamma|_{z=\zbar}~.
\ee

Having understood that 
the spectrum of open strings attached to a longitudinal 
D-brane is controlled not by the free parameter $\lambda$
appearing in (\ref{lambda}),
but by the definite quantity
$1/(2\nu^{2}\Gs^{2})$, where $\nu$ is the F-string number density
defined in (\ref{quant}), 
we can proceed with the review of \cite{go}.
As explained there, in view of (\ref{longbc}) it is natural to 
implement a `doubling trick', extending $\beta(z)$ and 
$\tbeta(\zbar)$ to the entire complex plane by setting,
for all $z$ in the upper-half plane,
\be \label{longdoubl}
\beta(\zbar)={1\over 4\nu^{2}\Gs^{2}} \pbar\tgamma(\zbar), \qquad
\tbeta(z)={1\over 4\nu^{2}\Gs^{2}} \p\gamma(z).
\ee
The boundary conditions (\ref{longbc}) then 
amount to requiring continuity of $\beta,\tbeta$ across the real axis. 
We thus have the mode expansions
\bea \label{longmodeexp}
\beta(z)= \sum^{\infty}_{n=-\infty}\beta_{n}z^{-n-1}
&\Longrightarrow&
\tgamma(\zbar)=4\nu^{2}\Gs^{2}\beta_{0} \log \zbar +\gamma_{0} 
- 4\nu^{2}\Gs^{2}\sum_{n\neq 0}{\beta_{n}\over n}\zbar^{-n},\\
\tbeta(\zbar)= \sum^{\infty}_{n=-\infty}\tbeta_{n}\zbar^{-n-1}
&\Longrightarrow&
\gamma(z)=4\nu^{2}\Gs^{2}\tbeta_{0} \log z +\gamma_{0} 
- 4\nu^{2}\Gs^{2}\sum_{n\neq 0}{\tbeta_{n}\over n}z^{-n}, \nonumber
\eea
the energy-momentum tensor 
$$
T^{\para}(z)=4\nu^{2}\Gs^{2}:\beta(z)\tbeta(z):~,
$$
and the Virasoro modes
\be \label{Lnlong}
L^{\para}_{n}=4\nu^{2}\Gs^{2}\sum_{l}:\beta_{l}\tbeta_{n-l}:~.
\ee
The non-zero commutators are 
\be \label{longcomm1}
[\gamma_{0},\beta_{0}]=1, \qquad 
[\tgamma_{0},\tbeta_{0}]=1, \qquad
[\tbeta_{n},\beta_{m}]={1\over 4\nu^{2}\Gs^{2}}n\delta_{n+m}~.
\ee
Using (\ref{pigamma}), we deduce that
\be \label{longppm}
p_{+}\equiv {1\over 2}(p_{0}+p_{1})= {i\beta_{0}\over\Ls}, \qquad
p_{-}\equiv {1\over 2}(-p_{0}+p_{1})= {i\tbeta_{0}\over\Ls}
\ee
Comparing the expansions for $\gamma=X^{+}/\Ls$ and  $\tgamma=X^{-}/\Ls$
in (\ref{longmodeexp}) 
with the standard open string mode expansion we see that 
$\Ls\gamma_{0}=x^{+}$, $\Ls\tgamma_{0}=x^{-}$, 
and it is also natural to define
\be \label{alphabeta}
\alpha^{-}_{m}=i2\sqrt{2}\nu^{2}\Gs^{2}\beta_{m}, \qquad
\alpha^{+}_{m}=i2\sqrt{2}\nu^{2}\Gs^{2}\tbeta_{m} \qquad 
\forall  \quad m\neq 0~.
\ee
We can then rewrite the commutators (\ref{longcomm1})
in the form
\bea \label{longcomm2}
{}&{}&[x^{+},p_{+}]=i=[x^{-},p_{-}], \\
{}&{}&[\alpha^{+}_{n},\alpha^{-}_{-n}]=2\nu^{2}\Gs^{2}n
                            =[\alpha^{-}_{n},\alpha^{+}_{-n}]
			    \qquad \forall \quad n>0~. \nonumber
\eea
Together with the contribution from the transverse part of the system, 
we thus clearly have the standard open string commutators,
\be \label{stdcommpm}
[x^{M},p^{N}]=iG^{MN}, \qquad
[\alpha^{M}_{m},\alpha^{N}_{n}]=G^{MN}m\delta_{m+n},
\ee
where $M,N=+,-,2,\ldots,D-1$, 
and we have introduced the Seiberg-Witten \cite{sw} open string metric  
$G_{+-}=1/(2\nu^{2}\Gs^{2})$, $G_{ij}=\delta_{ij}$.  

{}From (\ref{Lnlong}) we have
\be 
L^{\para}_{0}={4\nu^{2}\Gs^{2}}\Ls^{2}p_{+}p_{-}
  +{1\over 2\nu^{2}\Gs^{2}} \sum_{n=1}^{\infty}\alpha^{+}_{-n}\alpha^{-}_{n}
  +{1\over 2\nu^{2}\Gs^{2}} \sum_{n=1}^{\infty}\alpha^{+}_{-n}\alpha^{-}_{n}~,
\ee
which means the open string spectrum takes the usual form, 
\be
\Ls^{2}G^{MN}p_{M}p_{N}
   +\sum_{n=1}^{\infty}\alpha^{M}_{-n}\alpha^{N}_{n}G_{MN}=0~,
\ee
\emph{except that the metric is non-standard}. 
This mass-shell condition can be written 
out explicitly as
\be \label{p0long}
\nu^{2}\Gs^{2}
  \left[ (p_{0})^{2}-(p_{1})^{2}\right]- p_{\perp}^{2}= 
  {N_{osc}\over \Ls^{2}}~,
\ee
where $N_{osc}$ denotes the number operator for the oscillators.
The standard NCOS convention \cite{sst2,ncos} is to perform a 
$\nu$-dependent rescaling of the transverse 
closed string metric such that the factor 
$\nu^{2}\Gs^{2}$ is common to all terms in the left-hand side,  
and then absorb that factor 
through a redefinition of the effective string length, to be left with
\be \label{p0longNCOS}
(p_{0})^{2}-(p_{1})^{2}- k_{\perp}^{2}= 
  {N_{osc}\over \ape}~, \qquad 
\ape \equiv \nu^{2}\Gs^{2}\Ls^{2}~.
\ee
The point emphasized in \cite{wound} and reiterated here is that,
while these steps are sensible from the NCOS 
perspective, where one restricts attention to a particular D$p$-brane 
setup (with a specific value for $p$ and $\nu$), they are not convenient
when dealing with the complete Wound string theory,
where one should employ the same closed string 
metric and string length for all possible configurations.
Doing this one arrives at expressions like (\ref{p0long}), which
manifestly displays the dependence of the open string spectrum
on the relevant parameters.

Notice that, as advertised in Section \ref{unwoundgosec},
the open string spectrum
(\ref{p0long}) is independent of the parameter $\lambda$ appearing 
in (\ref{lambda}), so one is certainly free to set $\lambda$ to any 
desired value. 
The simplest choice is $\lambda=0$, which amounts to removing
the first term from 
the energy spectra (\ref{p0wound}) and (\ref{p0trans}).

\subsubsection{Transverse D-branes}
\label{transgosec}

Besides (\ref{longbc}), an `obvious' way to satisfy 
(\ref{gobc}) would be to demand that $\delta\gamma=0$ (and either 
$\delta\tgamma=0$ or $\tbeta=0$)  at $z=\zbar$. 
There is a problem with these boundary conditions, 
however: since $\gamma(z)$ is analytic, requiring it to be constant 
on the boundary forces it to be constant everywhere.
Physically, this seems counterintuitive, because it would
mean that the entire body of the string (and not just the endpoints) 
has a fixed position along $x^{+}$. Mathematically,
the problem is that these boundary conditions would be incompatible
with the algebra (\ref{comm}), because they require setting 
the \emph{creation} operators $\gamma_{n<0}$ to zero.
The way out is to
require $\gamma+\tgamma$ (and not $\gamma$ or $\tgamma$ separately)
to be constant on the real axis--- this is precisely as expected for 
a D-brane orthogonal to the $x^{1}$ direction.  
We then have $\delta\gamma=-\delta\tgamma$, or in 
other words $\p \gamma = -\pbar\tgamma$, at $z=\zbar$. 
To satisfy (\ref{gobc}) we must then set $\beta=-\tbeta$
at the boundary. Notice this implies that 
$T^{\para}(z)$ and $\tilde{T}^{\para}(\zbar)$ agree 
there (see (\ref{Tpara})), as needed for consistency. 
The mode expansions are 
\bea \label{transmodeexp}
\gamma(z)=-i\left(2wR\over\Ls\right)\log z 
     +\sum_{n}\gamma_{n}z^{-n}, \quad
\beta(z)=+\sum_{n} \beta_{n}z^{-n-1}, \\
\tgamma(\zbar)=+i\left(2wR\over\Ls\right)\log \zbar 
     -\sum_{n}\gamma_{n}\zbar^{-n}, \quad
\tbeta(\zbar)=-\sum_{n} \beta_{n}\zbar^{-n-1}. \nonumber
\eea
{}Using this in (\ref{pigamma}), 
we can extract the zero modes
\be \label{transp0p1}
p_{0}=i {\beta_{0}\over\Ls} 
   - {\lambda}\left({wR\over\Ls^{2}}\right), \qquad
p_{1}=0.
\ee
{}From (\ref{Tpara}) and (\ref{transmodeexp}) it follows that
\be \label{transL0para}
L^{\para}_{0}=-i\beta_{0}\left(2wR\over\Ls\right)+ N_{\para}~.
\ee
Together with (\ref{transp0p1}) and (\ref{transL0perp}), this 
implies that the condition $L_{0}=0$ is equivalent to
\be 
p_{0}={\lambda}{wR\over \Ls^{2}}+{\Ls^{2}p_{\perp}^{2}\over 2wR}
    +{N_{\para}+N_{\perp}\over 2wR}, 
\ee
which is in agreement with (\ref{p0trans}). 
The strings with `fractional winding' 
(extending between two different D-branes) analyzed in 
Section \ref{transsec}
can be accommodated here as well, simply by taking $w$ to 
be an arbitrary (positive) real number.
The treatment of the exceptional $w=0$ states is parallel
to the closed string case analyzed in Section \ref{unwoundgosec}.

\section{Supergravity Description}
\label{sugrasec}

We will now discuss some of 
the results of \cite{wound,go} and the previous sections
from the point of view of supergravity duals. 
For this purpose we should consider a system of 
$K\gg 1$ coincident longitudinal D$p$-branes in Wound string
theory. As explained in \cite{wound,om},
this is precisely the setup
known as $p+1$ NCOS theory \cite{sst2,ncos}. 
The supergravity duals for NCOS theories were worked out
in~\cite{ncos,harmark1,sahakian}; 
from the point of view of the parent string 
theory they are `near-horizon' limits of D-branes
with a worldvolume electric field.
We emphasize
here that in this case the duality is between a supergravity
background and a theory of strings (the NCOS/Wound
theory), instead of a field theory as in the usual case \cite{malda}. 

In the presence of the D-branes, and for a finite radius $R$ of the 
longitudinal direction,
Wound string theory contains not only open strings attached
to the branes, but also positively-wound 
closed strings which live in flat 
space \cite{km,wound,go}. 
In the decompactification limit $R\to\infty$, the worldvolume theory 
on the branes (still a full-fledged string theory) decouples from the 
closed strings.

Now, it is important to realize
that the `near-horizon' limit that defines the supergravity duals
of \cite{ncos,harmark1} is different from the familiar AdS/CFT scaling 
\cite{malda}. In particular, the NCOS limit (\ref{woundlim})
keeps transverse \emph{proper} 
distances fixed in units of $\ls$, whereas the Maldacena scaling
requires that $r\ll \ls$. It is therefore
interesting to ask exactly what objects 
are kept by the scaling in the supergravity description.

To address this question, consider the supergravity background dual
to $p+1$ NCOS \cite{harmark1},
\bea \label{sugrabg}
\frac{ds^2}{l_s^2} &=&
\frac{1}{L_s^2} H^{\frac{1}{2}}\frac{K^2}{G_s^2 \nu^2}
 \frac{u^{7-p}}{R_D^{7-p}} \left(-dX_0^2 + dX_1^2\right) +
\frac{1}{L_s^2} H^{-\frac{1}{2}}\left( dX_2^2+\ldots+dX_p^2\right)\\
&& +\frac{1}{L_s^2} H^{\frac{1}{2}}
    \left( dX_{p+1}^2+\ldots+dX_{d-1}^2\right) \\
g_{\mbox{\scriptsize eff}} &=& g_s e^{\phi} = H^{\frac{5-p}{4}}\frac{K}{\nu}
\frac{u^{\frac{7-p}{2}}}{R_D^{\frac{7-p}{2}}} \\
\frac{B_{01}}{l_s^2} &=& \frac{1}{G_s^2 L_s^2}\frac{K^2}{\nu^2}
\frac{u^{7-p}}{R_D^{7-p}}~,  \\
\eea
where
\bea
H &=& 1+ \frac{R_D^{7-p}}{u^{7-p}} \\
u^2 &=&  X_{p+1}^2+\ldots+X_{d-1}^2 \\
R_D^{7-p} &=& 
\frac{1}{7-p}\frac{K^2}{\nu}\frac{L_s^{7-p}}{V(\bS^{8-p})}~,
\eea
and $V(\bS^{8-p})$ is the volume of the $\bS^{8-p}$ sphere. Notice 
that we have rewritten the expressions of \cite{harmark1} 
in the units
and notation of the previous sections. In particular,
$X_1$ is regarded as being compactified on a circle of
radius $R$, and
$\nu$ is the
density of fundamental strings bound to the $K$ D$p$-branes,
as defined in (\ref{quant}).

It is easy to see that a signal propagating
outwards in this background at the speed of light, that is, obeying
\be
\frac{du}{dX_0} =
\frac{K}{G_s\nu}\frac{u^{\frac{7-p}{2}}}{R_D^{\frac{7-p}{2}}}~,
\ee
reaches the boundary $u=\infty$ in a finite time (for $p<5$),
whereas a
massive particle
will never be able to reach $u=\infty$. In this
sense the background (\ref{sugrabg})
behaves like a box, just like AdS space.

On the other hand, due to the presence of the $B$-field, 
some of the properties of the background (\ref{sugrabg}) are
certainly very different from those of AdS.
This difference can be seen most explicitly by considering
a fundamental string wound around the $X^{1}$ direction.
Besides the usual Nambu-Goto term, the relevant worldsheet
action includes of course a coupling to the $B$-field, and so reads
\bea \label{straction}
S &=& {1\over 2\pi l_s^2}\int d\tau d\sigma \
\left(\sqrt{h}-B_{01}\partial_\sigma X_1\right) \\
 &=& \int dX_0\, \frac{K^2}{\nu^2}\frac{R}{G_s^2L_s^2}
 \frac{u^{7-p}}{R_D^{7-p}} (H^{\frac{1}{2}} -1)~. \nonumber
\eea
In the second step we have assumed
a time-independent configuration,
and have
made the static gauge choice $X_0=\tau$, $X_1=\sigma R$.
{}From here we can infer that the string lives in a potential
\be
V(u) = \frac{K^2}{\nu^2}\frac{1}{G_s^2L_s^2}R \frac{u^{7-p}}{R_D^{7-p}}
\left(\sqrt{1+\frac{R_D^{7-p}}{u^{7-p}}}-1\right)~.
\ee
For $u\rightarrow 0$ we have $V(u)\rightarrow 0$, whereas
near the boundary
\be \label{asympot}
V(u) \simeq
\frac{K^2}{\nu^2}\frac{R}{G_s^2L_s^2}\left(\frac{1}{2} -\frac{1}{8}
\frac{R_D^{7-p}}{u^{7-p}} \ldots \right)
\ee
The second term gives an attractive Newtonian potential, whereas
the first
gives the energy necessary to pull the string out to infinity,
which is then
\be
\Delta E = \frac{1}{2} \frac{K^2}{\nu^2}\frac{R}{G_s^2L_s^2}~.
\label{DE}
\ee
The fact that the string needs
only a finite energy to 
reach the boundary is due to a cancellation between
the two terms in  (\ref{straction}), and so
depends crucially on the 
relative sign between them, which reflects a particular choice of 
orientation. The sign chosen in (\ref{straction}) is 
appropriate for a string which winds in the direction of the $B$-field.
The potential for an oppositely wound string would diverge at infinity.
This restriction to $w>0$ suggests 
that the string in question is the supergravity 
counterpart of a closed string in the Wound/NCOS theory.
Strings of this type  are known as long strings \cite{MMS,SW15,mo},
and have appeared before in other discussions of supergravity duals.

We can verify this picture quantitatively by noting 
that in the Wound/NCOS side
of the duality,
the expression (\ref{DE}) should give 
the energy necessary 
to pull one closed string out of the F1/D$p$ bound state.
The tension of a bound state of $K$
D$p$-branes and $\nu$ fundamental strings per unit $p-1$ volume
in the limit (\ref{woundlim}) is \cite{km,wound}
\be \label{woundDp}
\hat{T}_{K,\nu}= {K^2\over 2(2\pi)^{p}\nu\Gs^{2}\Ls^{p+1}}~,
\ee
so the total energy of the bound state is
\be
E_{K,\nu}
=\frac{1}{2}\frac{K^2}{\nu}\frac{R}{(G_sL_s)^2}
{r_{2}\ldots r_{p}\over L_{s}^{p-1}}~,
\ee
where $r_{2},\ldots,r_{p}$ denote the transverse radii.
When the number $N$ of fundamental strings in the bound state changes 
by one, 
the number density $\nu$ can be seen from (\ref{quant}) to 
change by $\delta\nu= \Ls^{p-1}/(r_{2}\ldots r_{p})$,
resulting in an energy change which is exactly the same as
in (\ref{DE}). 
We thus confirm
that the long strings living near the boundary of the supergravity
background correspond to the wound
closed strings of the dual picture. 
The usual short strings are essentially confined
to the region $u<R_{D}$; they correspond to the open strings living on
the brane, as in the familiar AdS/CFT mapping.\footnote{To be precise,
in the AdS/CFT correspondence only the massless modes of the open
strings remain in the limit, whereas here the whole tower of 
modes is kept.}
That the open/closed strings in NCOS theories have properties
analogous to those of the AdS$_{3}$
short/long strings discussed in \cite{mo} had been noted
already by Klebanov and Maldacena \cite{km}.

Further evidence for the identification of long strings with closed 
strings is given by the fact that for $u\gg R_D$ (i.e., 
close to the boundary), the
worldsheet action (\ref{straction}) for a long string 
reduces to the Gomis-Ooguri \cite{go}
action for a Wound string, Eq. (\ref{goaction}).
To prove this assertion, we can first note that 
in the region $u\gg R_D$, the background (\ref{sugrabg}) reduces to
\bea
\frac{ds^2}{l_s^2} &=& \frac{1}{L_s^2}\left[
\frac{K^2}{G_s^2\nu^2}\frac{u^{7-p}}{R_D^{7-p}}
\left(-dX_0^2 + dX_1^2\right)  +  dX_2^2 +\ldots + dX_{d-1}^2\right]~,\\
g_{\mbox{\scriptsize eff}} &=& g_s e^{\phi} = \frac{K}{\nu}
\frac{u^{\frac{7-p}{2}}}{R_D^{\frac{7-p}{2}}}~, \nonumber\\
\frac{B_{01}}{l_s^2} &=& \frac{1}{G_s^2 L_s^2}\frac{K^2}{\nu^2}
\frac{u^{7-p}}{R_D^{7-p}}~.  \nonumber
\eea
In Polyakov form,
the action (\ref{straction}) for a long string in this background
reads
\bea
S &=&\frac{1}{L_s^2} \int d\tau\,d\sigma \left[\frac{K^2}{G_s^2\nu^2}
\frac{u^{7-p}}{R_D^{7-p}} \left(
-\eta^{ab}\partial_a X_0 \partial_b X_0 +
\eta^{ab} \partial_a X_1 \partial_b X_1 +
2 \epsilon^{ab} \partial_a X_0 \partial_b X_1 \right)\right.  \nonumber\\
&& \qquad\qquad\qquad\left. + \eta^{ab} \left(
\partial_a X_2 \partial_b X_2 + \ldots +
\partial_a X_{d-1} \partial_b X_{d-1} \right)\right]~.
\eea
In the region $u\gg R_D$ we can define a small parameter
\be
\delta = \frac{G_s^2 \nu^2}{K^2} \frac{R_D^{7-p}}{u^{7-p}}~.
\ee
Upon making the identifications
$\ls^{2}= L_s^2 \delta$ for the string length, and 
\be
g_{\mbox{\scriptsize eff}} = \frac{K}{\nu} 
\frac{u^{\frac{7-p}{2}}}{R_D^{\frac{7-p}{2}}}
= G_s \frac{1}{\sqrt{\delta}}
\ee
for the string coupling,
it becomes clear
that the limit $u\to\infty$ is the same
as (\ref{woundlim}). 
Since this is the limit
that led to (\ref{goaction}), one can evidently use the 
Lagrange-multiplier trick of Gomis and Ooguri, to
obtain the $\beta$-$\gamma$ system as in~\cite{go}.
In a sense, this result is not surprising, since the limit 
(\ref{woundlim}) plays a role in the derivation of the 
supergravity background (\ref{sugrabg}).
The point which is worth emphasizing is that,
whereas short strings essentially see this
background as a box, the above result
shows that long strings see an asymptotically flat
space.

The above analysis
makes it rather clear that even though superficially
(from the point of view of the parent string theory) 
the limit (\ref{woundlim})
used to obtain the 
supergravity background (\ref{sugrabg})
appears to be a 
near-horizon limit, it is actually
quite different in nature from 
the Maldacena limit \cite{malda}.  The calculations of
\cite{ncos,harmark1} are best regarded 
 as a derivation of the macroscopic field 
configuration produced by a large number of longitudinal 
D-branes in Wound string theory.
{}From the point of view of the Wound theory, the duality 
in question \emph{does not involve any additional limits},
so it is simply based on the possibility of describing
the D-branes of this theory either 
by means of open strings or 
through the supergravity background they create.
The analogous duality
in a conventional string theory would be not the AdS/CFT 
correspondence, 
but the equivalence between the full asymptotically 
flat black brane backgrounds of \cite{hs}
and Polchinski's open-string description of
D-branes \cite{polchrr}. This equivalence is the starting point
of Maldacena's analysis \cite{malda}, 
and of the various recent attempts to
obtain a holographic dual for the full 
asymptotically flat backgrounds \cite{ghkk,gh,intri,d3holo,rvr}.

Given the results of Section \ref{unwoundsec},
we would expect the fluctuations of the background (\ref{sugrabg})
to include a special class of
unwound strings, corresponding to
the Newtonian gravitons of Wound string theory.
It is easy to see how they appear. Consider the metric
\be
\frac{ds^2}{l_s^2} = \eta_{\mu \nu} dx^{\mu} dx^{\nu}~,
\ee
where
\be
\eta_{+-}= \frac{1}{\ls^{2}}~, \qquad
\eta_{ij}= \delta_{ij}~. 
\ee
For a perturbation $h_{\mu \nu}$ around this background,
the Einstein equation reduces to
\be
R_{\mu \nu} = \frac{1}{2} \eta^{\delta\sigma}
\left(h_{\nu\sigma,\mu\delta}+h_{\mu\sigma,\nu\delta}-h_{\mu\nu,\delta\sigma
}-h_{\delta\sigma,\mu\nu}\right) = 0~,
\ee
from where it is clear
that terms containing $\eta^{+-} \sim l_s^2$
disappear in the limit $l_s^2\rightarrow 0$, allowing for more
solutions. Some of these solutions will not be physical, in the sense
that they will not correspond to any solution with $l_s^2\neq 0$. 
As we emphasized in Section 
\ref{unwoundgosec}, for the unwound states 
$l_s^2$
should be considered to be small but non-vanishing, so 
solutions of this type should be discarded. 
Solutions describing 
coordinate transformations are of course associated with null 
states. The remaining solutions can be seen to correspond to
the Newtonian gravitons discussed in Section \ref{unwoundsec}.

It is also interesting to ask how the results of the present
section generalize to the transverse D-branes of Wound string
theory. One can determine the supergravity background
that these branes generate
by starting with the standard D-brane solution 
in the parent string theory \cite{hs}, (with a superposed
$B_{01}$-field,) and then taking the limit
(\ref{woundlim})+(\ref{lambda}). As usual,
to comply with the requirement of
periodicity along the compact longitudinal dimension, one can 
either set up a periodic array of localized sources in the covering
space, or consider a longitudinally smeared source. Either way, one 
obtains a metric whose $00$ and $11$ components diverge everywhere 
relative to the others, in contrast
with the longitudinal D-brane case (\ref{sugrabg}), where they
only diverge asymptotically. 
This divergence can be dealt with using the methods of \cite{go},
and it indicates that, 
just like in the flat case (\ref{woundlim}),
only positively wound strings can propagate in this background.
We will elaborate on this set of issues elsewhere.

\section{Conclusions} \label{conclsec}

We have learned that, in a very 
specific sense, gravity is present in Wound string theory---
as discussed in Section \ref{polesec},
unwound strings with zero 
oscillator number, including a graviton, are part of the theory.
On-shell, these strings are 
forced to have vanishing transverse momentum, which as explained in 
Section \ref{zerosec} 
means that they are irrelevant as asymptotic states. 
Off-shell, their transverse momentum is arbitrary, and as the only 
massless states in the theory, they are the mediators of
all long-range interactions. 
As a result of the scaling of the metric in the limit (\ref{woundlim})
that defines the theory,
these messenger particles propagate at infinite speed,
and so the interactions they mediate
are instantaneous \cite{go}. 
This is an expression of the non-relativistic nature of
Wound string theory, apparent also in
the T-dual DLCQ description.

The presence of gravity in Wound string theory should not really
come as a surprise--- this is, after all, the reason why it has been
possible to discuss supergravity duals for the longitudinal
D-branes of this theory\footnote{In this connection 
we would like to emphasize
the point, made already in Section \ref{sugrasec}, that 
from the perspective of the Wound theory the duality in question is
analogous not to the AdS/CFT correspondence, but to Polchinski's 
identification of D-branes with R-R black branes.} 
\cite{ncos,harmark1}. As one would expect,
a collection of a large number of objects in the theory (wound
strings, D-branes or NS5-branes) does set up macroscopic
gravitational and Kalb-Ramond fields (and possibly others). The
point made clear by the results of 
\cite{go} and the present paper is that these fields are
Newtonian in character: they follow the source instantaneously.
So, even though gravity is present in the theory, it is still
true that there are no finite-time fluctuations of the
gravitational field--- there are no gravitons, in the traditional
sense of the word. Such spacetime fluctuations are at the root of
the traditional conceptual difficulties in attempts to understand
quantum gravity, so the hope remains that Wound string/NCOS
theory, devoid of such complicating features, could facilitate
our understanding of the underlying structure of string theory.

Note that the above statements about gravity in Wound string theory
can be generalized to all of the other 
Wrapped brane theories; e.g., Wrapped M2-brane theory 
\cite{wound,go} contains Newtonian 
gravitons which, on-shell, can carry momentum only along the 
two `longitudinal' directions (i.e., the directions along which the 
metric is not scaled to zero).

In Section \ref{transsec} we worked out
the excitation spectrum for transverse 
D-branes, and found that open strings with $w=0$ give rise to the 
expected gauge field and collective coordinates. 
The non-relativistic character of the theory is apparent here from the 
fact that there are no waves on these branes: the photons on their 
worldvolume, just like the gravitons in the bulk, are Newtonian, and 
the $p_{\perp}=0$ restriction on the scalars implies that the branes can 
only be translated rigidly. As in the case of gravity, 
a macroscopic source can set up a non-trivial 
gauge/scalar field configuration, 
which follows the source instantaneously.
It would be interesting, then, to look for
the analog of Born-Infeld strings \cite{calmal,gibbons}
in this context.

Besides explaining how all of 
the above conclusions follow from a direct 
examination of
the limit (\ref{woundlim}) that defines the theory,
we have shown how they can be extracted from
the worldsheet formalism developed by 
Gomis and Ooguri \cite{go}. As explained in Section 
\ref{unwoundgosec},
a direct application of the Lagrangian
of \cite{go} to the $w=0$ states
would lead to erroneous conclusions; special care must be taken to retain 
terms which would be subleading for $w>0$, but are necessary to project 
out negative-norm states in the case of unwound strings.
We have also verified in Section \ref{transgosec} that
the Gomis-Ooguri formalism
yields the correct spectrum for open 
strings ending on transverse D-branes.

Some additional remarks regarding the approach of 
\cite{go} have been made at various places. 
In Section \ref{unwoundgosec} 
it was pointed out that the Gomis-Ooguri action (\ref{goaction})
attains its simplest form if from the beginning 
we fine-tune the $B$-field to its critical value (i.e., set the free 
parameter $\lambda$ in (\ref{lambda}) equal to zero).
It was also shown there that the treatment of \cite{go} can easily 
be extended to the worldsheet fermions, resulting in
the action (\ref{psigoaction}), where $\psi^{0},\psi^{1}$ have been 
traded for a system of anticommuting ghosts.
Setting $B=1$ would appear to cause difficulties
when longitudinal D-branes are present, so in Section 
\ref{longgosec} we have retraced the steps of \cite{go} for that 
case to explain why there is in fact no problem.
The point is that, in addition to $B$,
for each longitudinal brane one must specify the value of the 
worldvolume electric field. When this is done, the 
open string spectrum is understood to be independent of the 
free parameter $\lambda$; its explicit dependence on the parameters 
$\Gs,\Ls$ of the Wound theory and the F-string density $\nu$ on the 
brane can be seen in (\ref{p0long}).  
In this connection we stress the point,
made already in \cite{wound}, that the 
NCOS conventions are not convenient when dealing with
the full Wound theory, because 
they obscure the dependence of physical quantities on the relevant 
parameters.  

Finally, in Section \ref{sugrasec} we considered the known
supergravity duals for Wound string theory in the presence of
longitudinal D-branes (i.e., NCOS theory) \cite{ncos,harmark1}. 
We observed in particular that the supergravity description 
accounts not only for the open strings  
attached to the branes, but also for
the wound strings which can move away from them. Whereas
the former are described by the usual local perturbations of 
the background (short strings), the latter are 
visible as long strings analogous to those of \cite{MMS,SW15,mo}.
That the open/closed strings in NCOS theories have properties
analogous to those of the AdS$_{3}$
short/long strings discussed in \cite{mo} had been noted
already by Klebanov and Maldacena \cite{km}.
Work is in progress regarding the extension of these ideas to
more general supergravity backgrounds, where one finds the 
interesting 
result that the presence of NS-NS or R-R fields
can cause a probe D-brane or F-string to become unstable at 
radii smaller than some critical value. 
We hope to report on this and related issues in the near future.

\vspace*{0.3cm}
\noindent
{\textbf{Note Added:}} 
While this paper was being written,
the three interesting works \cite{goteborg,sundell,hi2} 
appeared, which make 
remarks related to our Section \ref{sugrasec}.
In particular, our analysis of the potential for the long fundamental 
string is S-dual to
the discussion of a D1-brane probe in the supergravity 
background dual to NCYM \cite{hi1,mr} that is the subject of
Section 3.2 of \cite{hi2}.
{}From our perspective the latter background represents the 
fields set up by a large number of D3-branes in Wound D-string theory 
\cite{wound,go}, and the results of \cite{hi2} indicate that a wound 
D-string moving in this background sees an asymptotically flat space.

\section{Acknowledgements}

AG would like to thank Ansar Fayyazuddin,
Subir Mukho\-padhyay, Bo Sundborg, and Aleksandr Zheltukhin
for useful conversations. 
MK is grateful to Esko Keski-Vakkuri for valuable 
discussions.
The work of UD was supported by the Swedish Natural Science
Research Council (NFR) and the Royal Swedish Academy of Sciences. 
The work
of AG was supported by the NFR.

\end{document}